\documentclass[%
 reprint,
superscriptaddress,
preprintnumbers,
nofootinbib,
amsmath,amssymb,
prb,
]{revtex4-2}

\usepackage{graphicx} % Include figure files
\usepackage{dcolumn} % Align table columns on decimal point
\usepackage{bm} % bold math
\usepackage{hyperref} % add hypertext capabilities
\usepackage[mathlines]{lineno} % Enable numbering of text and display math
\usepackage{amsfonts} % math font
\usepackage[caption=false]{subfig}  % use for side-by-side figures
\usepackage[percent]{overpic} % For overlaying text on images
\usepackage{braket}
\usepackage{orcidlink}

\newcommand{\ii}{\mathrm{i}}%i for imaginary part
\newcommand{\dd}{\mathrm{d}}

\newcommand{\im}{\text{Im}\:}
\newcommand{\re}{\text{Re}\:}

\newcommand*{\tr}{\operatorname{tr}}

\begin{document}

\preprint{APS/123-QED}

%%%%%%%%%%%%%%%%%%%%%%%%%%%%%%%%%%%%%%%
%% title
%%%%%%%%%%%%%%%%%%%%%%%%%%%%%%%%%%%%%%% 
\title{Universal quench dynamics of lattice $q$ fermion Yukawa Sachdev-Ye-Kitaev model}% Force line breaks with \\
 
\author{Haixin Qiu
% \,\orcidlink{0000-0003-3443-6665}
}
\email{haixin.qiu@theorie.physik.uni-goettingen.de}
\affiliation{%
Institute for Theoretical Physics, Georg-August-Universität Göttingen,\\
Friedrich-Hund-Platz 1, 37077 Göttingen, Germany
}%

\author{Stefan Kehrein
% \,\orcidlink{0009-0005-1119-4124}
}%
\email{stefan.kehrein@theorie.physik.uni-goettingen.de}
\affiliation{%
Institute for Theoretical Physics, Georg-August-Universität Göttingen,\\
Friedrich-Hund-Platz 1, 37077 Göttingen, Germany
}%

\date{\today}% It is always \today, today,

%%%%%%%%%%%%%%%%%%%%%%%%%%%%%%%%%%%%%%%
%% abstract
%%%%%%%%%%%%%%%%%%%%%%%%%%%%%%%%%%%%%%% 
\begin{abstract}
  We study the quantum quench of the Yukawa Sachdev-Ye-Kitaev model and one of its lattice extensions with $q$ fermions and one boson. Several equilibrium properties are computed for general $q$ with different parameter scaling within the large-$N$ dynamical mean field scheme. The non-Fermi liquid quench dynamics are studied by integrating the Kadanoff-Baym equations for switching off the lattice term with constant Yukawa coupling or quenching to different final Yukawa couplings. The post-quench oscillations and relaxation dynamics are insensitive to the quench amplitudes deep inside the non-Fermi liquid phase. With weak lattice coupling quenches, we find universal thermalization dynamics similar to the SYK model; however, with two temperatures and two distinct relaxation rates for bosons and fermions, both signal Planckian relaxations without quasiparticles, as in strange metals.
\end{abstract}

\maketitle

% \tableofcontents

%%%%%%%%%%%%%%%%%%%%%%%%%%%%%%%%%%%%%%%
%% introduction
%%%%%%%%%%%%%%%%%%%%%%%%%%%%%%%%%%%%%%% 

\section{Introduction}
\label{sec:level1.1}

The Yukawa-Sachdev-Ye-Kitaev (YSYK) model has been introduced recently~\cite{Wang2020,Esterlis2019} as a model hosts both the non-Fermi liquid phase~\cite{Pan2021,Wang2020a,Valentinis2023,Kim2021,Cichutek2024} and the pairing phase~\cite{Esterlis2019}, features that also appear in high temperature superconductors. The similarity is pushed forward by several recent studies for the equilibrium properties of the lattice extension of this model, by considering either the two-dimensional Fermi surface~\cite{Esterlis2021, Guo2022, Li2024, Sachdev2023, Patel2023, Nikolaenko2023} or by computing the superfluid stiffness~\cite{Valentinis2023} with a random coupling lattice. In both cases, qualitative agreements were found with those observed in cuprates.

Closely related to the Sachdev-Ye-Kitaev (SYK) model~\cite{Sachdev2015, Maldacena2016}—in which maximal quantum chaos, holographic duality to black holes, and non-Fermi liquid behavior play central roles in understanding dynamics~\cite{Eberlein2017, Larzul2022, Louw2021, Louw2022}—the Yukawa SYK model and its extensions inherit many of these properties. They have also been studied in the contexts of quantum chaos~\cite{Davis2023}, holography~\cite{Inkof2022}, thermalization~\cite{Hosseinabadi2023}, and tunable non-Fermi liquid behavior~\cite{Bashan2024}.

Away from the equilibrium, the dynamics of YSYK models are of significant interest from both the fundamental and phenomenological perspectives. They may give insights into the light-driven unconventional superconductors and also may give us further understanding of the non-equilibrium processes in strange metals with interesting non-Fermi liquid behaviors, such as the "Planckian" transport.  Motivated by this, the quantum quenched YSYK model has been studied previously both for the superconducting state~\cite{Grunwald2024} in a closed system and the normal state~\cite{Hosseinabadi2023} with an external phonon bath coupled. However, the quench dynamics of isolated YSYK models remain largely unexplored, even in the case of the original YSYK model.

% \haixin{this work}
In this work, we study the equilibrium and non-equilibrium properties of a lattice extension of the YSYK model with $q$-fermion interaction. We mainly focus on half-filling where the chemical potential $\mu=0$. After the disorder average our model is effectively a SYK$_{q}$-YSYK$_{q}$ model, where $q$ is a positive even integer. Numerically, we integrate the corresponding Kadanoff-Baym equations of the large-$N$ saddle point dynamical mean field equations, starting from a correlated thermal initial state. In this setup, the lattice hopping $v$, the boson-fermion Yukawa coupling $g$, and the number of fermion pairs in the Yukawa interaction can be varied. 

While analytical large-$q$ solutions of the SYK model~\cite{Maldacena2016, Tarnopolsky2019} have been extensively studied and successfully applied, analytically controllable solutions for YSYK models are still lacking. Some finite-$q$ results may give insights into the large-$q$ solutions. From a broader point of view, the YSYK is also closely related to questions from Kondo physics~\cite{Parcollet1998, Haule2002}, the Holstein model~\cite{Murakami2015, Dee2019}, and the Migdal-Eliashberg theory~\cite{Wu2019, Yuzbashyan2022}. 

Although in real materials, driving with light will lead to much more involved light matter couplings than simple quantum quenches, to get insight into the non-equilibrium behaviors of non-Fermi liquids that without quasiparticles, e.g., for studying the linear in $T$ relaxations, quenches provide the simplest and one of the most tractable approaches. We present numerical results for the quench dynamics for the lattice YSYK model from different initial temperatures and several quench protocols. Although there are a lot of freedoms to choose quench protocols, we focus on two cases: quench $g$ with $v$ fixed and quench $v$ with $g$ fixed for simplicity. In the main text, the latter case, quench $v$ with $g$ fixed, will be discussed in detail because it shows clear Planckian dynamics.

% \haixin{Organization}
This work is organized as follows. 
In Sec.~\ref{sec:model_action} we present the model and effective action, the saddle point equation, and observables. We also give some numerical details. 
In Sec.~\ref{sec:equilibrium_properties} we start our analysis with the equilibrium properties of the YSYK$_q$ and lattice YSYK$_q$ model.  
In Sec.~\ref{sec:fix_g_quench} is our central result. We show non-equilibrium quench dynamics of our model when the Yukawa coupling $g$ is fixed while the lattice coupling $v$ varies during the quench. Here we fit the intermediate time temperature relaxation rates and find they are linear in the final temperature.
In Sec.~\ref{sec:more_about_universal_relaxation} we discuss possible explanations of those universal relaxation dynamics through where we further derive some kinetic equations.
The Sec.~\ref{sec:conclusion} is the conclusion. 
In Appendix~\ref{sec:fix_v_quench} we present the results for quenches with the lattice coupling $v$ fixed. Several detailed derivations are also collected in the Appendix~\ref{sec:finite_q_ysyk_real_time}, \ref{eq:more_on_equilibrium_properties}, \ref{sec:kinetic_eq_details}, \ref{sec:equal_time_eq_details} and \ref{sec:finite_q_ysyk_imaginary_time}.

\section{The model}
\label{sec:model_action}
Here we derive our $q$ fermion lattice YSYK model, where we will present the Lagrangians, effective actions, saddle point equations, and definitions of observables. We also give some details about numerics. More details on the disorder average can be found in Appendix~\ref{sec:finite_q_ysyk_real_time}. From now on we use $\hbar=k_{\mathrm{B}}=1$.

\subsection{Model}
The Lagrangian of the Yukawa Sachdev-Ye-Kitaev model with $q$ fermion boson interaction can be defined as $L= L_{0}+L_{g}+L_{v}$. The non-interacting part is given by
\begin{equation}
  \begin{aligned}
    L_0(t)=&
    \sum_{i}^{N}\sum_{\boldsymbol{x}\sigma}\psi_{i\sigma}^{\dagger}(t,\boldsymbol{x})(\ii\partial_{t}+\mu)\psi_{i\sigma}(t,\boldsymbol{x})\\
    &+ \sum_{k}^{M}\sum_{\boldsymbol{x}} \frac{1}{2} \phi_{k}(t,\boldsymbol{x}) (-\partial_{t}^{2} - \omega_0^2) \phi_{k}(t,\boldsymbol{x}) 
  \end{aligned}
\end{equation}
where $\phi$ represent scalar bosons with bare frequency $\omega_0$, and $\psi$ complex fermions with chemical potential $\mu$. Here $i,j$ are indices of $N$ complex fermion flavors, $\sigma=\uparrow,\downarrow$ for the spins and $k$ for $M$ boson flavors. The boson-fermion interaction part is
\begin{equation}
    L_{g}(t) =       \sum_{I J k \sigma \boldsymbol{x}}\frac{1}{\sqrt{M}} \left(\frac{1}{\sqrt{N}}\right)^{q-1} g_{I J k}(t) [\psi_{I\sigma}^{\dagger}\psi_{J\sigma}\phi_{k}](t,\boldsymbol{x})
\end{equation}
Here we defined compact indices $I = i_{1}\ldots i_{\frac{q}{2}}$ and $\psi_{J} = \psi_{j_1} \ldots \psi_{j_{\frac{q}{2}}}$.  The lattice coupling part is
\begin{equation}
    L_{v}(t) = \sum_{\braket{\boldsymbol{x}_1,\boldsymbol{x}_2}}\sum_{I J\sigma}^{N} \frac{v_{I J}(t,\boldsymbol{x}_1,\boldsymbol{x}_2)}{(\sqrt{N})^{q-1}}  \psi_{I\sigma}^{\dagger}(t,\boldsymbol{x}_1)\psi_{J\sigma}(t,\boldsymbol{x}_2)
\end{equation}
The $g_{IJk}$ and $v_{IJ}$ are random couplings. The $q$ is an even number.  We assume the Gaussian random couplings
\begin{equation}
  \label{eq:disorder_gaussian_complex}
  \begin{aligned}
   & \mathbf{E}[g_{IJ k}^{*} g_{LP l}] = g^{2}\frac{2}{q} \delta_{IL}\delta_{JP}\delta_{kl} \\
   & \mathbf{E}[v_{IJ k}(\boldsymbol{x}_1)^{*} v_{LP l}(\boldsymbol{x}_2)] = v^{2}\frac{2}{q} \delta_{IL}\delta_{JP}\delta_{kl}\delta_{\boldsymbol{x}_1,\boldsymbol{x}_2}
  \end{aligned}
\end{equation}
which gives only normal state solutions and excludes superconducting pairing states. Here, the $\mathbf{E}[\dots]$ means disorder average.

\subsection{Disorder averaged effective action and saddle point equations}

The local disorder averaged $G$-$\Sigma$-$D$-$\Pi$ effective action $S_{\mathrm{eff}} =S_{\mathrm{eff},x}$ is
\begin{equation}
  \begin{aligned}
     \frac{1}{N} &\ii S_{\mathrm{eff}} = \tr \log (G_{0}^{-1}-\Sigma) - \frac{\lambda}{2} \tr \log(D_0^{-1}-\Pi) \\
    & +   \frac{2}{q} \int_{t_1,t_2} v(t_1) v(t_2)   \left[G(t_1,t_2)G(t_2,t_1)\right]^{\frac{q}{2}}               \\
     & + \ii  \frac{2}{q}\int_{t_1,t_2} g(t_1)g(t_2) [G(t_1,t_2) G(t_2,t_1)]^{\frac{q}{2}} D(t_1,t_2)            \\
     & +  \int_{t_1,t_2} \Sigma(t_1,t_2) G(t_1,t_2) - \frac{\lambda}{2} \int_{t_1,t_2} \Pi(t_1,t_2) D(t_1,t_2).
  \end{aligned}
\end{equation}
Here $G_0^{-1}=\ii \partial_t+\mu$ and $D_0^{-1}=-\partial_t^2-\omega_0^2$, and the trace $\tr$ is taken over time variables. Note that the time integration contour in the effective action can be chosen as the three-branch Keldysh contour. Here, we suppressed spin and coordinate indices and defined
\begin{equation}
  \lambda=\frac{M}{N}
\end{equation}
The disorder-averaged partition function can now be written as
\begin{equation}
  Z = \int \dd[G,\Sigma,D,\Pi]
  \exp{\ii N S_{\mathrm{eff}}
  }
\end{equation}
If $N$ is large, the saddle point contributions are dominated and are exact for $N\to \infty$. They can be derived by functional variations over $G,\Sigma, D,\Pi$ dynamical mean fields. It is convenient to define
\begin{equation}
  R_{q-2}(t_1,t_2) \equiv [G(t_1,t_2)G(t_2,t_1)]^{\frac{q-2}{2}}
\end{equation}
With this, the saddle point equations are
\begin{equation}
  \begin{aligned}
    G(t_1,t_2)      & = \left[G_{0}^{-1}-\Sigma\right]^{-1}(t_1,t_2),      \\
    \Sigma(t_1,t_2) & =  (v(t_1)v(t_2) + g(t_1)g(t_2)\ii D(t_1,t_2)) \times\\
     & \qquad    R_{q-2}(t_1,t_2) G(t_1,t_2),                 \\
    D(t_1,t_2)      & =\left[D_{0}^{-1}-\Pi\right]^{-1}(t_1,t_2),          \\
    \Pi(t_1,t_2)    & = -\ii \frac{2 g(t_1)g(t_2)}{\lambda} \frac{2}{q} R_{q-2}(t_1,t_2) G(t_1,t_2) G(t_2,t_1).
  \end{aligned}
\end{equation}
See also~\eqref{eq:saddle_eq_0} in the appendix. If $q=2$ and $v=0$, the saddle point equations reduce to the usual YSYK ones~\cite{Wang2020, Esterlis2019}. For $q=2$ and non-zero $v$, it is the model considered recently in~\cite{Valentinis2023}. In this work, we usually set $\lambda=4/q^2$. The bare boson frequency is fixed by
\begin{equation}
  \omega_0=1
\end{equation}
From now on, it will serve as the energy unit.

\subsection{Definitions of observables}
Here we briefly define and discuss the observables we use to analyze the dynamics.

\paragraph{Occupation density}
Occupation densities of the fermion and boson are defined as equal-time functions
\begin{equation}
  n^{(f)}(t) = -\ii G^<(t,t),\quad n^{(b)}(t) = -\ii D^<(t,t)
\end{equation}
In numerics, those quantities are accessible for the full computed time range because they do not require Fourier transforms.

\paragraph{Wigner transform}
Here we use $G$ as an example to define the Wigner transform
\begin{equation}
  G(t_1,t_2) = G(t_{\mathrm{a}}+t_{\mathrm{r}}/2,t_{\mathrm{a}}-t_{\mathrm{r}}/2).
\end{equation}
The relative and average time are $t_{\mathrm{r}}=t_1-t_2$ and $t_{\mathrm{a}} = (t_1+t_2)/2$. One can then perform a Fourier transform on the relative time $t_{\mathrm{r}}\to\omega$ and write
\begin{equation}
  G(\omega,t_{\mathrm{a}}) = \int \dd t_{\mathrm{r}}  e^{\ii \omega t_{\mathrm{r}}} G(t_1,t_2)
\end{equation}
The same is defined for other two-point functions, e.g. $D(t_1,t_2)$, $\Sigma(t_1,t_2)$ and $\Pi(t_1,t_2)$. From two-point functions in Wigner coordinates, one can obtain time-dependent spectral functions and effective temperatures.

\paragraph{Spectral functions}
We define the time-dependent fermion spectral function as
\begin{equation}
  A^{(G)}(\omega,t_{\mathrm{a}}) = -2 \im G^R(\omega,t_{\mathrm{a}})
\end{equation}
and the boson spectral function as
\begin{equation}
  A^{(D)}(\omega,t_{\mathrm{a}}) = -2 \im D^R(\omega,t_{\mathrm{a}})
\end{equation}
and similarly for $A^{(\alpha)}$, where $\alpha \in \{\Sigma,\Pi\}$.

\paragraph{Distribution functions}
The time-dependent effective distributions $f^{(G)}$ and $b^{(D)}$ can be defined as
\begin{equation}
  1-2 f^{(G)} =  G^{K}(\omega,t_{\mathrm{a}})/(G^{R}(\omega,t_{\mathrm{a}})-G^{A}(\omega,t_{\mathrm{a}}))
\end{equation}
\begin{equation}
  1+2 b^{(D)} =  D^{K}(\omega,t_{\mathrm{a}})/(D^{R}(\omega,t_{\mathrm{a}})-D^{A}(\omega,t_{\mathrm{a}}))
\end{equation}
The superscripts indicate the Green's functions they are from. Note, it is also possible to define distribution functions from self-energies.

\paragraph{Effective temperature}
The effective distributions and the effective temperatures can be defined through Wigner Green's functions by assuming the distribution functions have the same form as those in equilibrium. For instance, the effective temperature extracted from the propagators can be written as~\cite{Hosseinabadi2023}
\begin{equation}
  \label{eq:def_beta_G}
  \tanh(\beta^{(G)}\omega/2) = G^{K}(\omega,t_{\mathrm{a}})/(G^{R}(\omega,t_{\mathrm{a}})-G^{A}(\omega,t_{\mathrm{a}}))
\end{equation}
and the boson effective temperature through
\begin{equation}
  \label{eq:def_beta_D}
  \tanh(\beta^{(D)}\omega/2) =  (D^{R}(\omega,t_{\mathrm{a}})-D^{A}(\omega,t_{\mathrm{a}}))/D^{K}(\omega,t_{\mathrm{a}})
\end{equation}
In numerics, the $\beta^{(G, D)}$ are extracted from numerical partial by using the frequency points $\omega=\pm 0.0167 \omega_0$.

\paragraph{Interaction energy}
The interaction energy~\cite{Hosseinabadi2023} for the YSYK model can be derived by adding a source that couples to the $H_{\mathrm{int}}$ then perform disorder average and $G$-$\Sigma$ trick, and finally setting $J=0$. By adding a source term to the action $S[v,g,J H_{\mathrm{int}}]=S[v,g]+\ii \int_{x} J(x)H_{\mathrm{int}}[v,g](x)$, the interaction energy becomes
\begin{equation}
  \braket{H_{\mathrm{int}}}=  \left[\frac{\delta}{\delta J(x)} \int \dd[\psi,\phi] \dd[v,g]\mathcal{P}[v,g] e^{\ii S[v,g,J H_{\mathrm{int}}]} \right]_{J=0}
\end{equation}
where $\mathcal{P}[v,g]$ is the classical disorder distribution of $v$ and $g$ couplings. The result is identical to the Galitskii-Migdal equation~\cite{Galitskii1958, Schueler2020}. This equation can be used in imaginary time or real-time computations.
The correlation energy is
\begin{equation}
  E_{\mathrm{corr}}^{(b)}(t)=\frac{1}{2} \im [\Pi \circ D]^{<}(t, t).
\end{equation}
Here, the $\circ$ is a convolution along the chosen time contour. For instance, the purely imaginary time version is supposed to be
$
  E_{\mathrm{corr}}^{(b)}=\frac{1}{2} \im [\Pi \circ D]^{<}(\tau=0)
$.
We have also checked that $E_{\mathrm{corr}}^{(b)}=-2 E_{\mathrm{corr}}^{(f)}$, the factor of $2$ is from the spin.

\paragraph{Relaxation time}
The relaxation time can be extracted from the effective temperatures or occupation numbers, etc. In many cases, those observables have damping behavior, and it may be convenient to assume oscillations on top of exponential relaxations. For a observable $O$, we define
\begin{equation}
  O(t) = C_{O} \cos(\omega_{O} t) e^{-\Gamma_{O} t}
\end{equation} 
where $C_{O}$ is a constant, $\omega_{O}$ is the damping oscillation frequency and $\Gamma_{O}$ is the relaxation rate. For temperature relaxation fittings, we usually set $\omega_{O}=0$.

\paragraph{Renormalized phonon frequency}
We define the renormalized boson frequency as
\begin{equation}
  \omega_{\mathrm{r}}^2 = \omega_0^2 + \re \Pi(\omega=0,t_{\mathrm{a}})
\end{equation}
which can be directly seen from the full boson propagator $D^{R}(\omega,t_{\mathrm{a}}) = [-\omega^2 -\omega_0^2-\Pi(\omega, t_{\mathrm{a}})]^{-1}$.

\begin{figure*}
  \centering
  \begin{overpic}[width=0.32\linewidth]{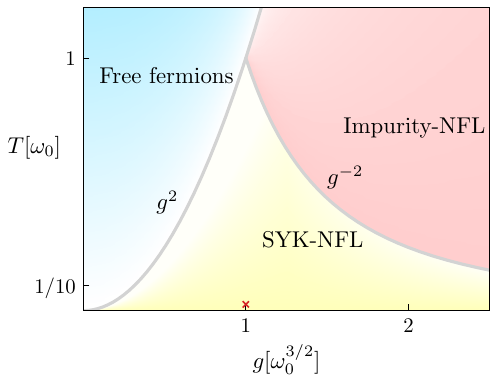}
    \put(7,69){(a)}
  \end{overpic}
  \begin{overpic}[width=0.32\linewidth]{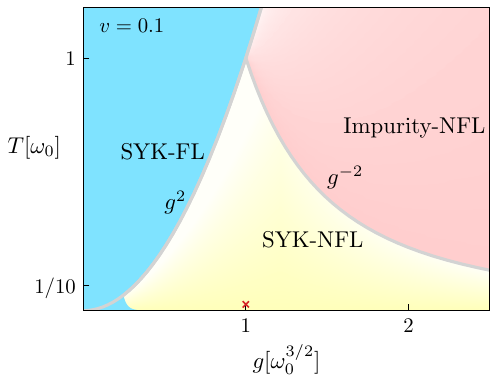}
    \put(7,69){(b)}
  \end{overpic}
  \begin{overpic}[width=0.32\linewidth]{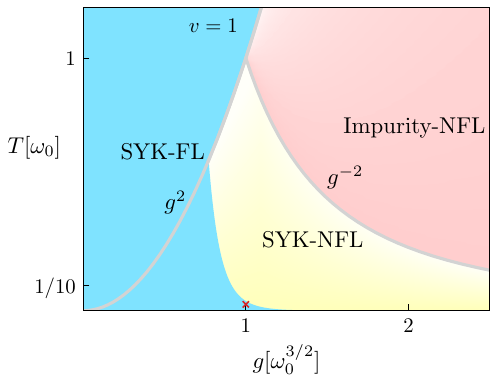}
    \put(7,69){(c)}
  \end{overpic}
  \caption{The half-filling phase diagram for $q=2$ according to the reference~\cite{Valentinis2023} using the equations provided there. (a) $v=0$. The two crossovers separate three phases: the free fermion phase, the non-Fermi liquid phase (SYK-NFL), and the impurity-like non-Fermi liquid phase (Impurity-NFL). (b) $v=0.1$. The Fermi liquid phase (SYK-FL) is marked blue. (c) $v=1$. The SYK-FL phase is larger than $v=0.1$. The red cross marks the typical pre-quench setting, $\beta=40$, $g=1$.
  }
  \label{fig:phase_diagram_q2}
\end{figure*}

\subsection{Numerical details}

The Kadanoff-Baym equations, with three-branch contour, are integrated by using codes based on \textit{The Non-Equilibrium Systems Simulation} (NESSi) package~\cite{Schueler2020}, which provides a well-tested framework. For time stepping, the fermion equations are integrated with order $5$ backward differentiation formula (BDF) as the predictor and Gregory integration for the Volterra integro-differential equation as the predictor. The correctors are iterations at the same time point until convergence. The boson equations are stepped through the Volterra integral equations into linear equations of order $5$. In most of the cases, we use imaginary time points $N_{\tau}=2000$ and real-time step $h=0.08$ with real-time points $N_{t} = 4000$. The quench time $t_0$ is usually set as the middle point of the full evolution period. Several cases are checked with $h=0.02$ without free evolutions before the quenches. 

The initial correlated Matsubara Green's functions for a given inverse temperature $\beta_i$ are computed first by a standalone self-consistent iteration code in imaginary time and Matsubara frequencies using fast Fourier transforms (FFT), with the imaginary time arrays over-padded to $3$ times the length of the frequencies. In most cases, we choose $\sim 4000$ Matsubara frequencies in a symmetric interval around zero. Note that the boson and fermion have different Matsubara frequency point positions due to statistics. The absolute tolerance is set to $10^{-7}$.  The resulting two-point functions are interpolated to $2000$ imaginary time points and sent to NESSi. The equilibrium problem is then solved again with the fixed-point method to the absolute tolerance $10^{-7}$ for fermion and boson propagators, which gives a double check of the initial state solutions. The Kadanoff-Baym equation is then evolved to quench time $t_0$ and after the quench to the final time $t_f$. We have also checked that the equal time observables, such as occupations and correlation energies, stay constant before the quench.

\section{Lattice YSYK-$q$ in Equilibrium}
\label{sec:equilibrium_properties}

Focusing on normal states, we first revisit several equilibrium properties of the lattice Yukawa Sachdev-Ye-Kitaev model. We also include results for finite-$q$. The equilibrium solutions are later used as initial conditions of the Kadanoff-Baym equations. 

\subsection{Phase diagram for $q=2$.}
The phase diagram of $q=2$ lattice YSYK has been previously studied~\cite{Esterlis2019, Valentinis2023}, where we show it in Fig.~\ref{fig:phase_diagram_q2} (a-c) in a temperature-coupling, i.e., $T-g$, diagram. Note that due to our choice of the complex disorder couplings, no superconducting pairing phase can appear, see~\eqref{eq:disorder_gaussian_complex}. If $v=0$, as in Fig.~\ref{fig:phase_diagram_q2} (a), two cross-overlines are separating three phases, and we call them, following the Ref.~\cite{Esterlis2019, Valentinis2023}, the Free fermion phase, the SYK  non-Fermi liquid (SYK-NFL) phase of the YSYK model, and the impurity-like non-Fermi liquid phase (impurity-NFL). The crossover lines are qualitatively $\sim g^{-2}$ and $g^{2}$, shown in light grey solid lines in Fig.~\ref{fig:phase_diagram_q2}. Switch on the lattice coupling $v$, as those in Fig.~\ref{fig:phase_diagram_q2} (b) and (c), we have the free fermion phase been replaced by a SYK Fermi liquid (SYK-FL) phase~\cite{Valentinis2023}. Those SYK-FL phases are colored blue in Fig.~\ref{fig:phase_diagram_q2} (b) and (c). One can see that for larger $v$ the area of the SYK-FL phases becomes larger. We can see that with $g=1$ and $T<1/10$ and $v=0$, the system is deep in the SYK-NFL phase. This is indeed the phase we will mainly consider.

\subsection{Finite $q$ two point functions in imaginary time}

If the system is deep inside the SYK-NFL phase, e.g., with low temperatures $T<1/10$ and $g\sim 1$, the long-time parts of the imaginary time Green's functions can be approximated well by conformal solutions. Solving the saddle point equations numerically by using Matsubara formalism, we can also get fermion and boson Green's functions, $G$ and $D$. The long (imaginary) time conformal solutions~\cite{Esterlis2019, Wang2020} are
\begin{equation}
  \label{eq:G_scaling_main}
  G_{\mathrm{c}}(\tau) \sim |\tau|^{\frac{-4\Delta}{q}} \overset{q=2}{\simeq}  |\tau|^{-0.84}
\end{equation}
\begin{equation}
  \label{eq:D_scaling_main}
  D_{\mathrm{c}}(\tau) \sim |\tau|^{-(2-4\Delta)} \overset{q=2}{\simeq} |\tau|^{-0.32}
\end{equation}
as for $q=2$, $\Delta \simeq 0.420374$. For finite $q$ other than $2$ the scaling exponent $\Delta$ can be determined through the equation
\begin{equation}
  \label{eq:finite_q_exponent_main}
  1 = \frac{- 4 \tan (2 \pi  \Delta ) (q-4 \Delta ) \tan \left(\frac{2 \pi  \Delta }{q}\right)}{(4 \Delta -1) \lambda  q^2}
\end{equation}
One can find more details about this equation in the Appendix~\ref{sec:q_dependent_exponents}.

In Fig.~\ref{fig:eq_G_D_vs_conformal} we show solutions from the saddle point equations in imaginary time and Matsubara frequencies numerically for different $q$ with $\beta=160 \omega_0^{-1}$. The scaling exponents with $\lambda=\frac{M}{N}=\frac{4}{q^2}$ are computed and listed in Table~\ref{tab:exponent_table1}. From the figures, we can see that in an energy scale where $\omega$ is much smaller than $T$, the numerically computed Green's functions follow the expected scaling behaviors.
\begin{figure}
  \centering
  \begin{overpic}[width=0.8\linewidth]{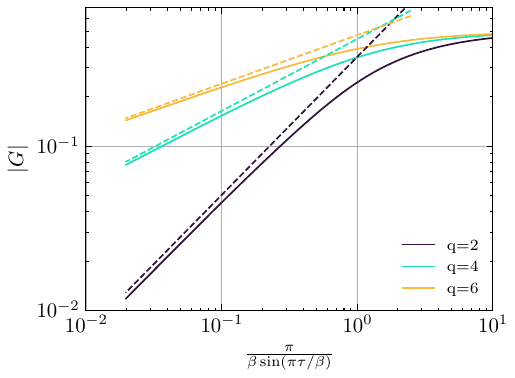}
    \put(22,65){(a)} 
  \end{overpic}
  \begin{overpic}[width=0.8\linewidth]{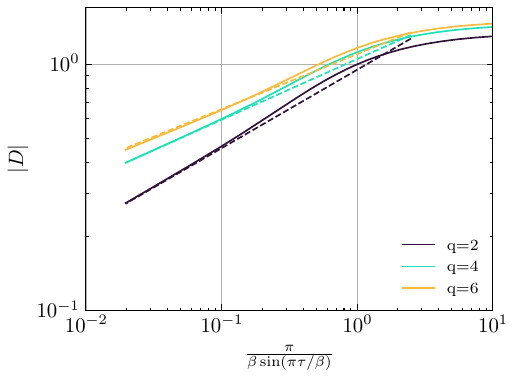}
    \put(22,65){(b)} 
  \end{overpic}
  \caption{Imaginary time Green's functions, (a) for fermion and (b) for boson. Solid lines are numerics and dashed lines are conformal solutions with the scaling dimensions in Table~\ref{tab:exponent_table1}. Here $\beta=160\omega_0^{-1}$, $g^2=\omega_0^3 2^{q-1}/q$, $\lambda=4/q^2$ and $v=0$. One can see the computed $q$ dependent exponents fit well. Note for reaching low temperature, we used $10^4$ Matsubara frequencies. One expects the agreement when $\frac{\pi}{\beta \sin(\pi \tau/\beta)}$ is small.}
  \label{fig:eq_G_D_vs_conformal}
\end{figure}

\begin{table}
\caption{Scaling exponent $\Delta$ for different $q$ with $\lambda=\frac{4}{q^2}$, from solving the \eqref{eq:finite_q_exponent_main}}
\label{tab:exponent_table1}
\begin{tabular}{|c|c|c|c|c|c|c|}
\hline
$q$      & 2        & 4        & 6        & 8        & 10             \\ \hline
$\Delta$ & 0.420374& 0.438401& 0.444108& 0.446943& 0.448643  \\ \hline
\end{tabular}
\end{table}
% 0.420374, 0.438401, 0.444108, 0.446943, 0.448643

\section{Quench with fixed Yukawa coupling $g$}
\label{sec:fix_g_quench}

Here we study the SYK interaction quench
\begin{equation}
v \to 0,  \quad g \to g
\end{equation}
The quench is realized through a Heaviside theta $v_i+\theta(t_0)\delta v$ where $t_0$ is the quench time. For finite $q$ other than $2$, we use the scaling
\begin{equation}
  \label{eq:g_v_lam_q_parameter_scaling}
  g_{q|i,f} = g\sqrt{\frac{2^{q-1}}{q}},  \quad v_{q|i,f} = v\sqrt{\frac{2^{q-1}}{q}},\quad \lambda_{q} = \lambda \frac{4}{q^2}
\end{equation}
to get a more uniform $q$ dependence. Note that by using these scalings, the $q$ dependent coupling is fully determined by the values of $g,v,\lambda$, which simplify the numbers in our discussions. Note that this also means the quench parameters are scaled as well.

\subsection{Spectral function and distribution}

\begin{figure*}
  \centering
  \begin{overpic}[width=0.37\linewidth]{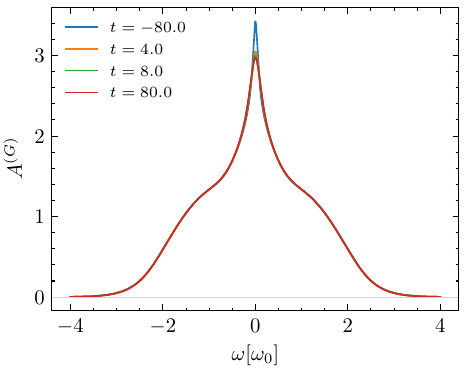}
    \put(1,69){(a)}
  \end{overpic}
  \begin{overpic}[width=0.4\linewidth]{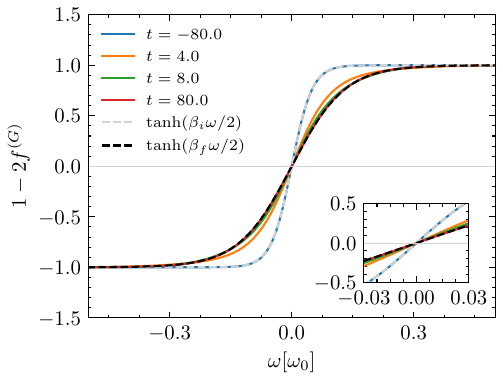}
    \put(1,65){(b)}
  \end{overpic}
  \begin{overpic}[width=0.4\linewidth]{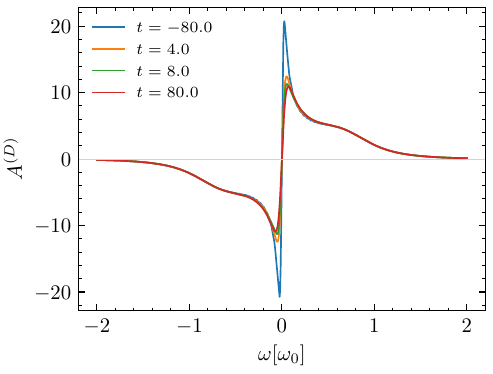}
    \put(1,65){(c)}
  \end{overpic}
  \begin{overpic}[width=0.4\linewidth]{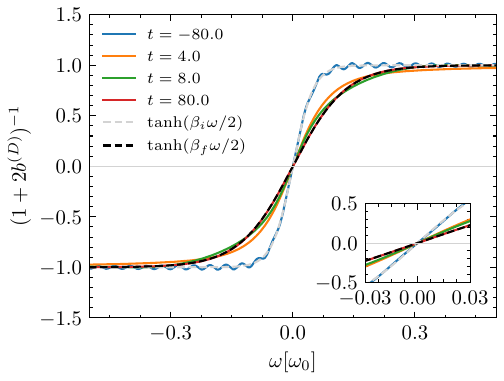}
    \put(1,65){(d)}
  \end{overpic}
  \caption{Spectral functions (a) for fermions and (c) for bosons. Effective distributions (b) for fermions and (d) for bosons. Here we fixed $g=1$, $q=2$ with quench protocol $v_i=0.1$, $v_f=0$. We use $\beta=40$ here. The insets of (c) and (d) are zoomed-in pictures. The spectral function changes mainly in the low-frequency region for this small quench. The distribution functions have smooth time dependence and can give well-defined effective temperatures that can be extracted from the slopes of effective distributions around $\omega=0$ and be used for the analysis of temperature dynamics. Here, the $\beta_i$ is the initial temperature, and the $\beta_f$ is the fit temperature at the latest time. One can see that the initial and final distributions match the expected equilibrium distribution.}
  \label{fig:fixg_q2_spectral}
\end{figure*}

In Fig.~\ref{fig:fixg_q2_spectral} we show spectral functions of fermions, $A^{(G)}(\omega,t_{\mathrm{a}})$ and bosons, $A^{(D)}(\omega,t_{\mathrm{a}})$, and their corresponding effective distributions for several centers of mass time $t_{\mathrm{a}}$, where $g=1$. From Fig.~\ref{fig:fixg_q2_spectral} (a) and (c), the shape of the spectral function agrees with that in previous studies~\cite{Valentinis2023}. One can see that for the small quench of the $v$, only the low-frequency part of the spectral functions reduces their height; this applies similarly to the fermions and bosons. The spectral functions also converge to their final value rather rapidly. In Fig.~\ref{fig:fixg_q2_spectral} (b) and (d), we show the effective distribution functions extracted from the corresponding two-point functions through the fluctuation-dissipation relation. For visibility we show the quantities $1-2 f^{(G)}(\omega,t_{\mathrm{a}})$ and $(1+2b^{(G)}(\omega,t_{\mathrm{a}}))^{-1}$, since they should be equal to $\tanh \frac{\beta\omega}{2}$ in equilibrium, thus well behaved. We can see from the figures that distributions are smooth curves without sharp peak structures and with a fast relaxation to their final values. In the insets of Fig.~\ref{fig:fixg_q2_spectral} (b) and (d) we show the zoomed-in low-frequency part of the distributions, where one can see that they are straight curves and give well-defined effective inverse temperatures $\beta_{\mathrm{eff}}$ through their slopes around $\omega = 0$. We will use the effective temperature extracted from those slopes to study the non-equilibrium temperature relaxations later in this section.

\subsection{Dynamics of boson occupations, renormalized frequencies, and correlation energies}

\begin{figure*}
  \centering
  \begin{overpic}[width=0.4\linewidth]{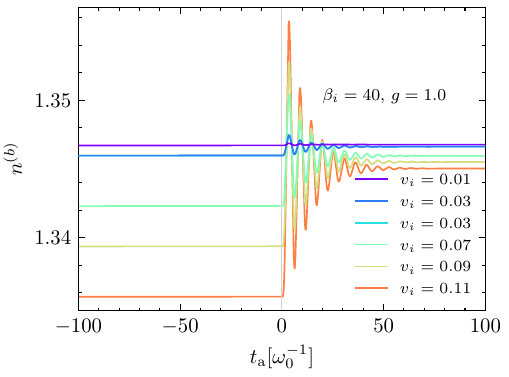}
    \put(0,65){(a)}
  \end{overpic}
  \begin{overpic}[width=0.41\linewidth]{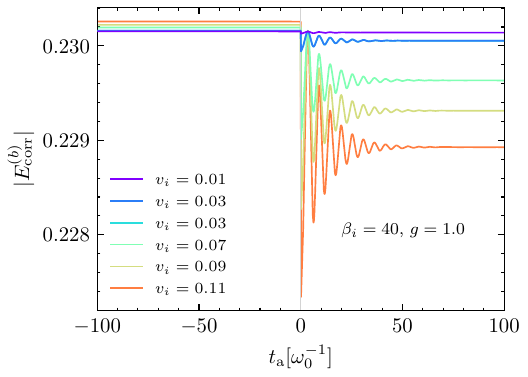}
    \put(0,65){(b)}
  \end{overpic}
  \begin{overpic}[width=0.4\linewidth]{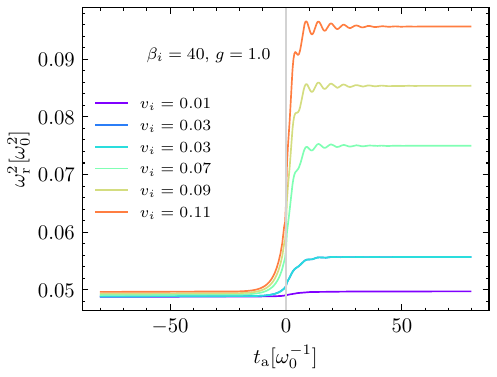}
    \put(0,65){(c)}
  \end{overpic}
  \begin{overpic}[width=0.41\linewidth]{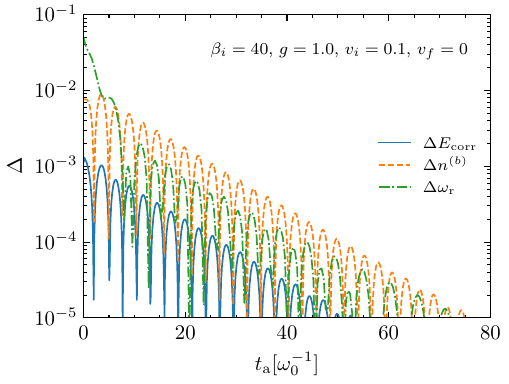}
    \put(0,65){(d)}
  \end{overpic}
  \caption{ Time dependences of boson occupations $n^{(b)}(t)$, boson correlation energies $|E^{(b)}_{\mathrm{corr}}(t)|$ and boson renormalized frequencies $\omega_{\mathrm{r}}^2(t_{\mathrm{a}})$. The protocol is $v_i$ to $v_f=0$ with fixed $g=1$, $q=2$. The lower right figure shows the log-log plot for $v_i=0.1$ of those three quantities, and the agreement of their time dependence can be directly observed. For $n^{(b)}(t)$ and $|E^{(b)}_{\mathrm{corr}}(t)|$, one can see that they stay constant before the quench at $t=0$. The precursors of the $\omega_{\mathrm{r}}$ before the quenches are because of using the center of mass slices two-point functions, which can contain the post-quenched part even when $t_{\mathrm{a}}<0$. We find damping oscillation dynamics are different from the temperature dynamics.
  }
  \label{fig:fixg_q2_occupation}
\end{figure*}

In Fig.~\ref{fig:fixg_q2_occupation} we show boson occupation $n^{(b)}$, the absolute value of boson correlation energy $|E^{(b)}_{\mathrm{corr}}|$, the squared boson renormalized frequencies $\omega_{\mathrm{r}}^2$ and a log-log plot of those three quantities. The overall picture is that their post-quench behavior is almost identical, with similar frequencies and relaxation rates. This can be best seen from Fig.~\ref{fig:fixg_q2_occupation} (d) where both decay rate and the oscillation periods of $n^{(b)}$, $|E^{(b)}_{\mathrm{corr}}|$ and $\omega_{\mathrm{r}}^2$ match each other. From Fig.~\ref{fig:fixg_q2_occupation} (a), we can see that for different initial $v$, the occupation oscillations share the same frequency. This can be understood~\cite{Murakami2015} by assuming that the boson field oscillates as $\phi(t)\sim \cos(\frac{1}{2}\omega_{\phi^2} t)$ where $\frac{1}{2}\omega_{\phi^2}$ is the oscillation frequency, then occupation is then $\braket{\phi(t)\phi(t)}$ oscillates with the frequency $\omega_{\phi^2}$. The relaxation rates are robust against different quench parameters. This is also true for quench protocols where $v$ is fixed and $g$ is varying, see Appendix~\ref{sec:fix_v_quench}. The other two quantities are shown in Fig.~\ref{fig:fixg_q2_occupation} (b) and (c) give similar dynamics. The Fig.~\ref{fig:fixg_q2_occupation} (c) shows that for larger quenches the system is driven farther from the quantum criticality since the critical point is the soft boson limit $\omega_{\mathrm{r}} \to 0$. This is consistent with the quench heating the system. We will see in the next section that compared with the $n^{(b)}$, $|E^{(b)}_{\mathrm{corr}}|$ and $\omega_{\mathrm{r}}^2$, the effective temperature dynamics behave differently.

\subsection{Time dependent effective temperature and linear in $T_f$ relaxation rate}

\begin{figure*}
  \centering
  \begin{overpic}[width=0.4\linewidth]{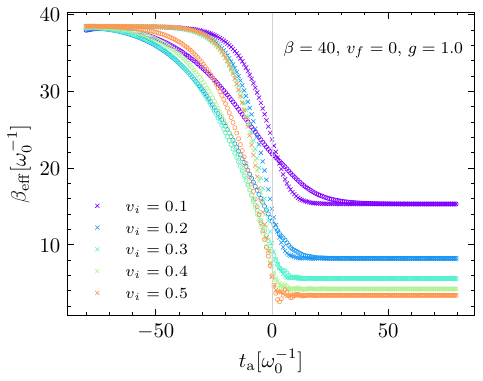}
    \put(1,70){(a)}
  \end{overpic}
  \begin{overpic}[width=0.4\linewidth]{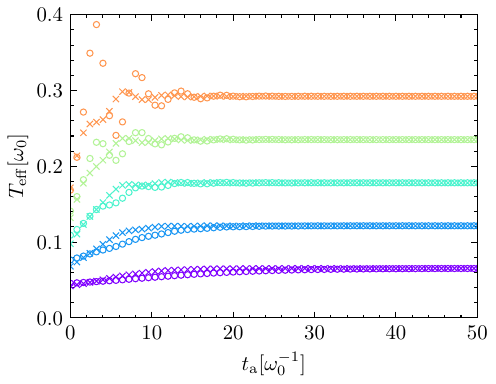}
    \put(1,70){(b)}
  \end{overpic}
  \caption{Center of the mass time dependence of effective inverse temperature and temperature.  We fixed $g=1$, $q=2$. The crosses are for fermions and circles for bosons. Note here we use strong quenches $v_i=0.1,0.2,0.3,0.4,0.5$ to give a broader view of the quench parameter region.  The data shown is sparse for visibility. From (a), we can see precursors of the effective inverse temperatures decaying. This is because the center of mass, i.e., Wigner coordinate, two-point function, is used for temperature extraction, which means even for $t_{\mathrm{a}}<0$, there can be post-quenched effects. For $t_{\mathrm{a}}>0$, the two-point functions are purely post-quenched. The effective temperatures are shown in (b). We can see that the bosons and fermions have different relaxation rates, and the quench affects the different subsystems differently.
  }
  \label{fig:fixg_q2_beta}
\end{figure*}

In Fig.~\ref{fig:fixg_q2_beta} we first show the time center of mass time dependence of effective temperature for bosons and fermions for five different initial $v_i$ and $v_f=0$. Note that the strongest quench we show in this plot is $v_i=0.5$, which is for a broader view of the effect of the quench parameter amplitudes and their resulting temperature relaxations. One can see from Fig.~\ref{fig:fixg_q2_beta} (a) that for $v_i=0.1$ the boson and fermion effective inverse temperatures $\beta_{\mathrm{eff}}$ exponentially relax to their final values, however the intermediate values and relaxation rates are distinct to each other. This is true for other quench parameters, however, the difference is that for larger quench, there are temperature oscillations, which make the analysis harder. There are again precursors of the center of mass time $t_\mathrm{a}$, which is again because the $t_\mathrm{a}$ slices cross both the $t_1$ and $t_2$ axes and thus can contain post-quench information when $t_{\mathrm{a}}<0$. Here we always quench at $t_1=t_2=0$ and the fermion Green's function $G(t_1,t_2)$. In Fig.~\ref{fig:fixg_q2_beta} (b) we plot the effective temperatures $T_{\mathrm{eff}}$ for the same set of initial $v$ and only show the post-quench part. It is clear that for large quench, e.g., $v_i=0.5$, the temperature oscillations are pronounced and the oscillation amplitudes are different for boson and fermion. Also, the effective temperature relaxations are faster when quench amplitudes are larger.

\begin{figure*}
  \centering
  \begin{overpic}[width=0.32\linewidth]{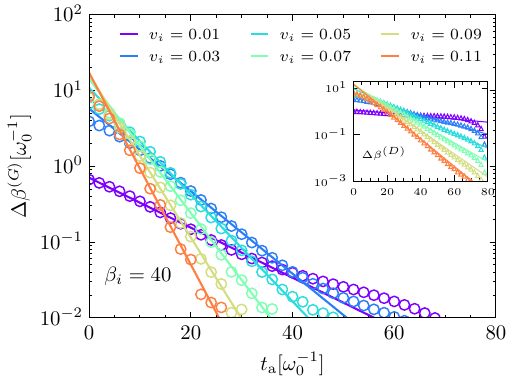}
    \put(1,70){(a)}
  \end{overpic}
  \begin{overpic}[width=0.32\linewidth]{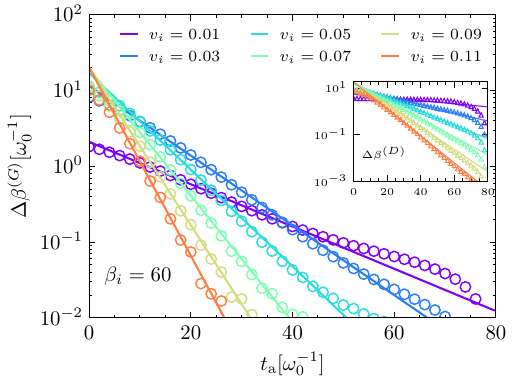}
    \put(1,70){(b)}
  \end{overpic}
  % \vskip\baselineskip
  \begin{overpic}[width=0.32\linewidth]{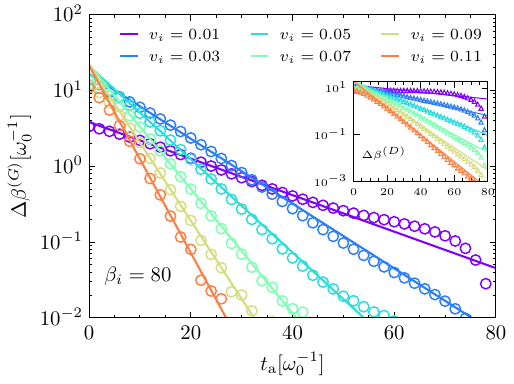}
    \put(1,70){(c)}
  \end{overpic}
  \caption{Temperature relaxations for $q=2$, $g=1$ and different initial temperatures. Note $\Delta \beta(t) = \beta(t)-\beta_f$. Here, markers are data, and solid lines are fitted curves or the guide of the eye. From (a-c), we can see that the exponentials fit well in a long intermediate time window. The effective inverse temperature relaxation rates $\Gamma_{\beta}^{(G,D)}$ are extracted from the propagators $G$ or $D$. The fitting results of this figure are summarized in Fig~\ref{fig:fixg_q246_Gamma} (a). 
  }
  \label{fig:fixg_q2_Gamma}
\end{figure*}

For a closer look at the temperature relaxations, we consider the small quenches $v_i<0.12$ and fit the intermediate time inverse temperatures to the exponential functions. In Fig.~\ref{fig:fixg_q2_Gamma} (a-c) we show the logarithmic scale plot, for three different initial temperatures $\beta_i=40,60,80$, of the fermion (boson in insets) effective inverse temperature differences
\begin{equation}
  \Delta \beta^{(G,D)}(t_{\mathrm{a}}) = \beta^{(G,D)}(t_{\mathrm{a}}) - \beta_f  
\end{equation}
where $\beta_f$ is the inverse temperature of the latest time in the computation, typically $\beta_f^{(G)} =\beta_f^{(G)} = \beta(t_{\mathrm{a}}=80/\omega_0)$. The $\beta^{(G)}$ and $\beta^{(D)}$ are defined previously in equations \eqref{eq:def_beta_G} and \eqref{eq:def_beta_D}. In Fig.~\ref{fig:fixg_q2_Gamma} (a-c), we show sparser data than the actual computed data for visibility. The solid lines in Fig.~\ref{fig:fixg_q2_Gamma} (a-c) are fitted curves where we assume the exponential decay of the inverse temperature
\begin{equation} 
  \Delta \beta^{(G,D)}(t_{\mathrm{a}}) \sim e^{-\Gamma_{\beta}^{(G,D)} t_{\mathrm{a}}}
\end{equation}
From Fig.~\ref{fig:fixg_q2_Gamma} (a-c), we can see that apart from the very long time behaviors, which may be from the numerical errors, the effective inverse temperatures of fermions show two stages of dynamics: a high-frequency non-universal change at short times and a long-time exponential decay. For small $v$, we find good agreement of the fits, whereas for larger $v$, there are some oscillations, especially for $\Delta \beta^{(G)}$.

The extracted fermion (boson in inset) inverse temperature relaxation rates versus final temperature $T_{f}$ are shown in Fig.~\ref{fig:fixg_q246_Gamma} (a), both fermions and bosons give linear dependence of the final temperature $T_f$. For the pure SYK$_{4}$ the linear dependence of the $T_f$ is observed in Ref.~\cite{Eberlein2017}. Here, the crucial difference is that we consider two species of particles, namely bosons and fermions, where both show strongly coupled relaxation behaviors, however, with distinct relaxation rates. 

In Fig.~\ref{fig:fixg_q246_Gamma} (a), to estimate the uncertainties of the relaxation rates, we do linear regressions of the $\log \Delta \beta(t)$ with lower cutoffs for $\Delta\beta$ at $10^{-1}$, $10^{-2}$ and $10^{-3}$, and get fitting results $\Gamma_1$, $\Gamma_2$, $\Gamma_3$.  Then the upper bounds of the error bars are $\max(\Gamma_1,\Gamma_2,\Gamma_3, \Gamma)$ and lower bounds are $\min(\Gamma_1,\Gamma_2,\Gamma_3, \Gamma)$. Note that in this way, the error bars do not reflect a normal distribution, since the data points for a given cutoff are highly correlated. The error bars should be understood as a qualitative estimation of the fitting uncertainties and cross checks of the nonlinear fittings. Bosons' data are from the same methods as fermions, except that due to slower relaxation the starting time is $t_{\mathrm{a}}=20$ and linear regression cutoffs are $10^{-1}$, $10^{-2}$, $10^{-3}$. We also checked that the fitting standard deviations are much smaller than the estimated uncertainties. 

\subsection{Temperature relaxations for $q\geq 4$ or $g\neq 1$}

To further clarify the origin of the linear $T_f$ dependence of the temperature relaxation rates, we further numerically evaluate two more setups: $q=2$ and with different fixed $g$, and  $q=4,6$ with $g=1$.

\begin{figure*}
  \centering
  \begin{overpic}[width=0.32\linewidth]{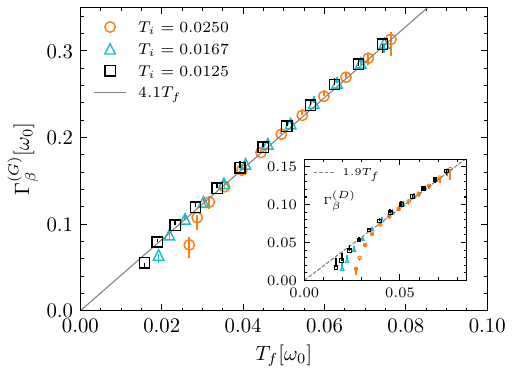}
    \put(1,70){(a)} 
  \end{overpic} 
  \begin{overpic}[width=0.32\linewidth]{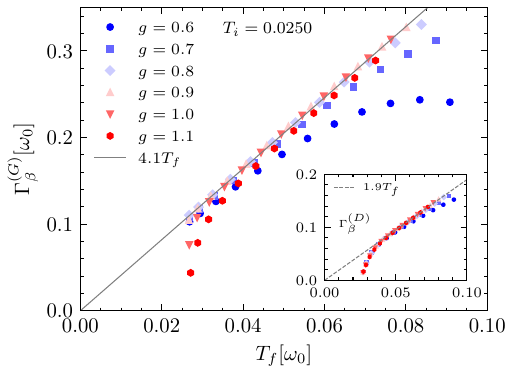}
    \put(1,70){(b)}
  \end{overpic}
  \begin{overpic}[width=0.32\linewidth]{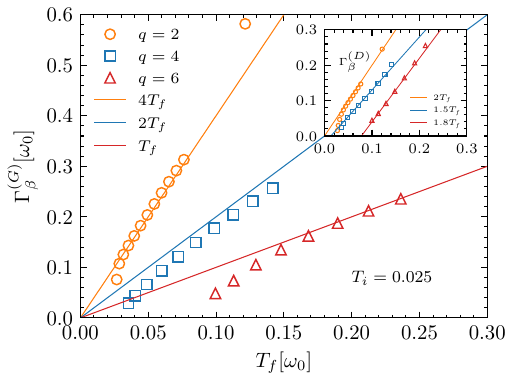}
    \put(1,70){(c)}
  \end{overpic}

  \caption{Temperature relaxations by quenching from finite $v$ to $0$. 
  (a) $v$ quenches with different initial temperatures $T_i=1/40,1/60,1/80$, $g=1$. 
  % We can see that after a small cut-off temperature, which is about the initial equilibrium temperature, the relaxation rates of fermions and bosons both show linear dependence on the final temperature $T_f$. The slopes of fermion and boson rates are different by about a factor of two. 
  For fermions, the data are taken after $t_{\mathrm{a}}=10$. Center values $\Gamma$ are from non-linear fits to exponential functions. The bars are from linear regressions with different lower cutoffs; see the main text for explanations.
  (b) Temperature relaxations for fermions and bosons (inset) with different constant $g$. The $T_i=1/40$. For increasing $g$, the curves converge to be linear in $T_f$ and share a single slope of about $4.1$. 
  (c) Temperature relaxations for fermions and bosons (inset) with different $q$ where $q=2,4,6$, $g=1$. Note $g^2/\omega_0^3$ is dimensionless, and $\omega_0=1$. Here $\beta_i=40$, i.e. $T_i=0.025$. Note the actual couplings for different $q$ are scaled as described in equation~\eqref{eq:g_v_lam_q_parameter_scaling}. The solid lines are the guide for the eye. We can see that the relaxation rates follow the tendencies of scaling dimensions, where for fermions $\sim \Delta/q$ and for bosons $\sim \Delta$. For each $q>2$, the linear dependencies can still be resolved. The top orange data in (c) is that with initial $v=0.2$ and has a strong effective temperature oscillation. Otherwise, the $q=2$ data in (c) are the same as that in (a). 
  }
  \label{fig:fixg_q246_Gamma}
\end{figure*}

In Fig.~\ref{fig:fixg_q246_Gamma} (b), we show the relaxation rates for different Yukawa couplings $g$ from $0.6$ to $1.1$. The overall tendency is clear: for higher $g$ the curves converge to a single slope $ \sim 4.1 T_{f}$ and show linear dependencies. The low $T_f$ parts of the larger $g$ curves have deviations from the linear behavior, especially for the points close to the initial temperature $T=0.025$. For the curves with $g\leq 0.8$, deviations from the line $4.1T_f$ become pronounced. This is because the post-quenched parameters are closer to the free fermion phase, thus giving much slower relaxations, i.e., smaller $\Gamma$, and are pulled downwards from the bound linear in $T_f$.

We also computed the relaxation rates for $q=4,6$. The $q > 6$ cases require a lower temperature to get deep inside the non-Fermi liquid phase, which is numerically not as feasible as the $q\leq 6$ cases. In Fig.~\ref{fig:fixg_q246_Gamma} (c), we show the temperature relaxation rates for different $q$, $q=2,4,6$, using the parameter scaling~\eqref{eq:g_v_lam_q_parameter_scaling} and the saddle point equations with $q \geq 2$. The extraction technique is the same as that used in Fig.~\ref{fig:fixg_q2_Gamma} and Fig.~\ref{fig:fixg_q246_Gamma} (a). For the $q=4,6$, we have $\Gamma_{\beta}^{((G))}\sim 1/q$ and $\Gamma_{\beta}^{(D)}\sim 1$ which is following the scaling dimension of the propagators \eqref{eq:G_scaling_main} and \eqref{eq:D_scaling_main}. This is expected since the post-quench Hamiltonian is a $q$ fermion YSYK without lattice coupling. For $g$ quenches with finite lattice coupling $v$, we evaluated the temperature relaxation, see Fig.~\ref{fig:fixv_q2_v0_Gamma} in Appendix~\ref{sec:fix_v_quench}, where the linear dependence is less pronounced, which can be a result of the competition of the lattice term.

\section{More about universal relaxations}
\label{sec:more_about_universal_relaxation}

Here we continue discussions about the universal quench dynamics or relaxations further. The hope is that for some time region, the approximated kinetic equations, i.e., approximations of the Kadanoff-Baym equations, may give some partial analytical insights. We will see that the equations remain rather complicated, and it is challenging to find analytical solutions.

\subsection{Temperature relaxation}

Inspired by the solutions obtained from the Kadanoff-Baym equation, we may approximate the intermediate time dynamics by assuming time local equilibrium and using time-dependent effective temperatures. The exponential growth of the effective temperature may be written as 
\begin{equation}
  \partial_{t} \left(T^{(f)}(t)-T_f^{(f)} \right) = \Gamma_{T}^{(f)} \left(T^{(f)}(t)-T_f^{(f)}\right)  
\end{equation}
\begin{equation}
  \partial_{t} \left(T^{(b)}(t)-T_f^{(b)} \right) = \Gamma_{T}^{(b)} \left(T^{(b)}(t)-T_f^{(b)}\right)  
\end{equation}
where we used $T^{(f,b)} \equiv T^{(f,b)}_{\mathrm{eff}}$. The relaxation rate is expected to be $\Gamma_{T}^{(f,b)} = 1/\Gamma_{\beta}^{(G,D)}$. From the evolution of the spectral functions and effective distributions, which look quasi-static, we expect $\Gamma_{T}$ may have some relation with the generalized collision integral and the quantum Boltzmann equations. However, we will see that it is rather difficult to derive precise analytical relations.

\subsection{About temperature relaxation: Approximate kinetic equations with scale separation}

For temperature relaxations, we may try to get differential equations of effective distributions. One way to achieve this is to apply the scale separation assumption. After Wigner transform and expanding to the lowest order of the gradient expansion~\cite{Kamenev2011, Aoki2014, Rammer1986,Abrikosov1975}, we get generalized quantum Boltzmann equations~\cite{Kamenev2011, Picano2021, Larzul2022} without assuming quasi-particles, see also Appendix~\ref{sec:kinetic_eq_details}. The resulting equations are
\begin{equation}
  -\ii \partial_{t_{\mathrm{a}}} F(\omega,t_{\mathrm{a}}) = I_{\mathrm{col}}^{(f)}[F,B]
\end{equation}
\begin{equation}
  -2\ii \omega \partial_{t_{\mathrm{a}}} B(\omega,t_{\mathrm{a}}) = I_{\mathrm{col}}^{(b)}[F,B]
\end{equation}
where $F$ and $B$ are related to distribution functions. The collision integrals under the same approximations are
\begin{equation}
  I_{\mathrm{col}}^{(f)} = \Sigma^{K}(\omega,t_{\mathrm{a}})-F(\omega,t_{\mathrm{a}})(\Sigma^R(\omega,t_{\mathrm{a}})-\Sigma^A(\omega,t_{\mathrm{a}}))
\end{equation}
\begin{equation}
  I_{\mathrm{col}}^{(b)} = \Pi^{K}(\omega,t_{\mathrm{a}}) - B(\omega,t_{\mathrm{a}})(\Pi^R(\omega,t_{\mathrm{a}})-\Pi^A(\omega,t_{\mathrm{a}}))
\end{equation}
We can also write Green's functions into time-dependent distributions and spectral functions under the scale separation approximation~\cite{Aoki2014, Kamenev2011}
$
  G^K(\omega,t_{\mathrm{a}}) = F(G^R-G^A) = F(\omega,t_{\mathrm{a}}) A^{(G)}(\omega,t_{\mathrm{a}})
$
and 
$
  D^K(\omega,t_{\mathrm{a}}) = B(D^R-D^A) = B(\omega,t_{\mathrm{a}}) A^{(D)}(\omega,t_{\mathrm{a}})
$
which gives the greater and lesser functions immediately in the form
\begin{equation}
  G^{\gtrless}(\omega,t_{\mathrm{a}}) = -\ii A^{(G)}\frac{F \pm 1}{2},
\quad
  D^{\gtrless}(\omega,t_{\mathrm{a}}) = -\ii A^{(D)}\frac{B\pm 1}{2}
\end{equation}
where we suppressed some variables for short. One can easily identify that $F=1-2f$ and $B=1+2b$, which are exactly the definitions of $F$ and $B$.

% \subsubsection{$q=2$}
We can write out the collision integrals with the help of $\Sigma^{R}-\Sigma^{A} = \Sigma^>-\Sigma^< $. Defining $[A \bullet  B](\nu) = \int_{\omega} A(\nu-\omega) B(\omega)$  the results are
\begin{equation}
  \label{eq:Ifcoll_q2}
  \begin{aligned}
    \frac{-2\ii}{g^2} I_{\mathrm{col}}^{(f)} =  
      &F ( (A^{(D)} B) \bullet  A^{(G)}+A^{(D)} \bullet ( A^{(G)} F) )\\
      &- (A^{(D)} B) \bullet ( A^{(G)} F) - A^{(D)} \bullet  A^{(G)}
  \end{aligned}
\end{equation}
\begin{equation}
  \label{eq:Ibcoll_q2}
  \begin{aligned}
  \frac{-\lambda \ii}{g^2} I_{\mathrm{col}}^{(b)} = 
  &F ( ( A^{(G)} F) \bullet  A^{(G)}_{-} -A^{(G)} \bullet ( A^{(G)}_{-} F_{-} ) ) \\
  &+ ( A^{(G)} F) \bullet ( A^{(G)}_{-} F_{-} ) -  A^{(G)} \bullet A^{(G)}_{-} 
\end{aligned}
\end{equation}
here $A_{-} = A(-\omega,t_{\mathrm{a}})$ and similarly for $F_{-}$. See also \eqref{eq:Ifcoll_q_equal_2} and \eqref{eq:Ibcoll_q_equal_2} in the Appendix. From the above collision integrals, we can find immediately that for $q=2$ the lattice coupling $v$ does not contribute to the kinetic equation. This is understandable because when $q=2$, the $v$ terms are quadratic and collisionless. For larger $q$, we also computed the collision integrals, which can be found from the Appendix, the equations \eqref{eq:Ifcoll_q_greater_2} and \eqref{eq:Ibcoll_q_greater_2}. The $v$ terms now appear in the collision integrals and are expected to give SYK dynamics, which is competing with the YSYK part. One can see that those equations are rather involved and difficult to solve analytically. Since we already have the results from the Kadanoff-Baym equations, we will not try to solve those kinetic equations numerically here.

\subsection{Occupation relaxation: equal-time equations}
\label{sec:occupation_equal_time_equations}

Here we discuss more about the boson occupation $n^{(b)}(t)$ oscillation with the frequency $\omega_{\phi^2}$ and decay rate $\Gamma_{\phi^2}$. The relevant equation will be the equal-time version of the Kadanoff-Baym equations, which can be found mostly in the literature about the generalized Kadanoff-Baym ansatz as an approximation. Here we will not further apply any approximations but only use the general exact equations, and we will focus on the equations for bosons.

We start with the Kadanoff-Baym equations
\begin{equation}
  [-\partial_{t_1}^2-\omega_0^2] D^<(t_1,t_2) = K_1(t_1,t_2)
\end{equation}
\begin{equation}
  [-\partial_{t_2}^2-\omega_0^2] D^<(t_1,t_2) = K_2(t_1,t_2)
\end{equation}
where the $K_1$ and $K_2$ are terms from usual Langreth rules, for which the full equations can be found in equations \eqref{eq:boson_kbe_1} and \eqref{eq:boson_kbe_2}. To get the equal time equations, some operations are needed to convert the second derivatives to first-order ones by factorizing the products and inverting one derivative to the right-hand side. We show how we do this in Appendix~\ref{sec:equal_time_eq_details}. The equal time equation can then be written as
\begin{equation}
  -\partial_{t} n^{(b)}(t) = C_{\mathrm{col}}^{(b)}(t)  
\end{equation}
The right-hand side convoluted collision integral is defined as
\begin{equation}
  C_{\mathrm{col}}^{(b)}(t) = [P_{+} \circ K_1](t,t)-[K_2 \circ M_{+}](t,t)
\end{equation}
where $P_{+}$ is a retarded function and $M_{+}$ is an advanced function, defined by
\begin{equation}
  P_{+}(t_1,t_2) = - \ii \theta(t_1-t_2)e^{+ \ii\omega_0(t_1-t_2)}
\end{equation}
\begin{equation}
  M_{+}(t_1,t_2) =  \ii \theta(t_2-t_1)e^{+ \ii\omega_0(t_1-t_2)}
\end{equation}
The simplest check is that $C_{\mathrm{col}}(t)=0$ if we have free collisionless bosons $K_1=0$ and $K_2=0$. 

One can also see by taking a step further that with suitable choices of $K_1(t_1,t_2)$ and $K_2(t_1,t_2)$, the above equations can give $n^{(b)}$ oscillations with the frequency $\omega_0$. Plugging in the $P_{+}$ and $M_{+}$ to $C_{\mathrm{col}}^{(b)}(t)$, we get
\begin{equation}
C_{\mathrm{col}}^{(b)}(t) = -\ii e^{\ii\omega_0}\alpha_1(t)-\ii e^{-\ii \omega_0 t} \alpha_2(t)
\end{equation}
with $\alpha_1(t) \equiv \int_{-\infty}^{t}\dd t_2 e^{-\ii \omega_0 t_2} K_1(t_2,t)$ and $\alpha_2(t) \equiv \int_{t}^{\infty}\dd t_2  K_2(t,t_2)e^{\ii \omega_0 t_2}$.  Now if we can have $\alpha_1(t) \sim \alpha_2(t)=\alpha$, which may be achieved by assuming $K_1$ and $K_2$ are peaked around $t$, we get $-\partial_{t} n^{(b)}(t) = -2\ii \alpha \cos (\omega_0 t) $ which has the solution $n^{(b)}(t) \sim \alpha \sin(\omega_0)/\omega_0$ up to boundary terms and prefactors. Of course, the actual solutions of the damped oscillations should be found by computing the $C_{\mathrm{col}}^{(b)}(t)$, and it will give frequency corrections and decay rates. We leave this analysis to future works.

\section{Discussion and conclusion}
\label{sec:conclusion}

In this work, we investigated the equilibrium two-point functions and quantum quench dynamics of the finite-$q$ extension of the normal state of the YSYK model and one of its lattice extensions. In contrast to previous studies~\cite{Hosseinabadi2023}, no external bath is coupled to the system, the relaxation dynamics are entirely governed by the intrinsic interactions of the YSYK model. The time evolutions are obtained by numerically integrating Kadanoff-Baym equations of large-$N$ saddle point equations, without introducing additional approximations. We analyzed spectral functions, effective distribution functions, effective temperatures, occupation densities, and other observables. Additionally, we derived simplified forms of the Kadanoff-Baym equations to obtain analytical insights into the relaxation behaviors.

Interestingly, as shown in Fig.~\ref{fig:fixg_q246_Gamma}, quenching the lattice coupling $v$ to zero with small amplitude reveals the emergence of two distinct transient temperatures associated with the bosons and fermions, respectively. For all considered values of $q = 2, 4, 6$ and initial temperatures $T_{i}=1/40,1/60,1/80$, the inverse temperature relaxation rates $\Gamma$ exhibit a linear dependence on the final temperature $T_f$, with fermions and bosons relaxing at different rates. This behavior bears similarities to that observed in SYK models~\cite{Eberlein2017}, but differs due to the interplay between bosonic and fermionic degrees of freedom.

The observed linear-in-$T$ relaxation rates reflect the absence of quasiparticles and are indicative of Planckian transport. In such systems, the relaxation rate $\Gamma$, defined as the inverse relaxation time $\tau_{\mathrm{relax}}$, scales linearly with temperature~\cite{Hartnoll2022, Patel2019}:
\begin{equation}
  \Gamma = \frac{1}{\tau_{\mathrm{relax}}} = f \frac{k_{\mathrm{B}} T}{\hbar},
\end{equation}
where the dimensionless prefactor $f$ is expected to be of order unity, $f\sim O(1)$. In the YSYK model, we find $f_{\mathrm{F}} \sim 4.1$ for fermions and $f_{\mathrm{B}} \sim 1.9$ for bosons, consistent with previous results for extended lattice SYK models, where $f \simeq 4.8$~\cite{Patel2019}, and for the SYK$_4$ model, where $f \simeq 2$~\cite{Eberlein2017}.

In summary, we have identified clear signatures of Planckian behavior in the lattice $q$-fermion Yukawa Sachdev-Ye-Kitaev model by investigating its non-equilibrium dynamics following quenches in the lattice couplings. The linear in $T_f$ relaxation spans over approximately a decade in final temperature $T_f$, with fermions and bosons exhibiting distinct relaxation rates, highlighting the system's intricate dynamics without quasiparticles.

%%%%%%%%%%%%%%%%%%%%%%%%%%%%%%%%%%%%%%%
%% acknowledgments
%%%%%%%%%%%%%%%%%%%%%%%%%%%%%%%%%%%%%%% 

\begin{acknowledgments}
We thank Rishabh Jha, Jan Louw, Vibhu Mishra, and Anastasia Enckell for useful discussions. This work is supported by the Deutsche Forschungsgemeinschaft (DFG) 217133147/SFB 1073 (Project No. B03) and 499180199/FOR 5522 (Project No. T1). This work used the Scientific Compute Cluster at GWDG, the joint data center of the Max Planck Society for the Advancement of Science (MPG) and the University of Göttingen, with partial funding from the Deutsche Forschungsgemeinschaft (DFG, German Research Foundation 405797229). 
\end{acknowledgments}

%%%%%%%%%%%%%%%%%%%%%%%%%%%%%%%%%%%%%%%
%% appendix
%%%%%%%%%%%%%%%%%%%%%%%%%%%%%%%%%%%%%%%
\appendix

\section{Disorder average finite $q$ YSYK in Real time}
\label{sec:finite_q_ysyk_real_time}
Now we consider real-time equations. Since the $v_{IJ}(\boldsymbol{x}$ gives a delta function in position space, the Lagrangian in the main text is equivalent to the following Hamiltonian in the sense of large-$N$ disorder averaged effective action
\begin{widetext}
\begin{equation}
  \begin{aligned}
    H_{\mathrm{int}} =& \left(\frac{1}{\sqrt{N}}\right)^{q-1}\sum_{\sigma}\sum_{i_1 i_2 \ldots i_{\frac{q}{2}}}\sum_{j_1 j_2\ldots j_{\frac{q}{2}}} v_{i_1\ldots i_{\frac{q}{2}} j_1 \ldots j_{\frac{q}{2}}}\psi^{\dagger}_{i_1 \sigma}\ldots\psi_{i_{\frac{q}{2}} \sigma}^{\dagger}\psi_{j_1 \sigma}\ldots\psi_{j_{\frac{q}{2}} \sigma}\\
    &+\left(\frac{1}{\sqrt{N}}\right)^{q-1}\frac{1}{\sqrt{ M}}\sum_{\sigma k} \sum_{i_1 i_2 \ldots i_{\frac{q}{2}}}\sum_{j_1 j_2\ldots j_{\frac{q}{2}}} g_{i_1\ldots i_{\frac{q}{2}} j_1 \ldots j_{\frac{q}{2}} k}\psi^{\dagger}_{i_1 \sigma}\ldots\psi_{i_{\frac{q}{2}} \sigma}^{\dagger}\psi_{j_1 \sigma}\ldots\psi_{j_{\frac{q}{2}}  \sigma} \phi_{k}
  \end{aligned}
\end{equation}
\end{widetext}
We may focus on the first line and integrate out $v$. Note from hermicity we have $v_{i_1\ldots i_{\frac{q}{2}} j_1 \ldots j_{\frac{q}{2}}}=v^{*}_{ j_{\frac{q}{2} \ldots j_1 } i_{\frac{q}{2}}\ldots i_1}$, and we can choose disorder potential to satisfy
\begin{equation}
  \mathbf{E}[v_{IJ}^{*}v_{KL}]=v^2 \frac{2}{q} \delta_{IK} \delta_{JL}
\end{equation}
where we defined collected index $I = i_1\ldots i_{\frac{q}{2}}$. The disorder average can be performed as
\begin{equation}
  \begin{aligned}
     & \mathbf{E}\left[ e^{\ii v_{IJ} O_{IJ}  }\right]
    = \exp\left\{\ii\left(\ii v^2 \frac{2}{q}\sum_{I J} O_{\bar{J}\bar{I}} O_{IJ }\right)\right\}
  \end{aligned}
\end{equation}
The $\bar{I} = i_{\frac{q}{2}}\ldots i_1$ is the reversed ordered compact index, the same for $\bar{J}$.
Note we always assume replica symmetry here. From the above considerations, the $v$ coupling part interacting action under a functional integral of fermionic fields $\overline{\psi}$ and $\psi$ can be written as
\begin{equation}
  \begin{aligned}
   \ii  &S_{\mathrm{int},v,\sigma} =\\
     &\ii^2 \frac{1}{N^{q-1}} v^2 \frac{2}{q}\int_{t_1 t_2} \sum_{IJ} \overline{\psi}_{\bar{J}\sigma}(t_1)\psi_{\bar{I}\sigma}(t_1)\overline{\psi}_{I\sigma}(t_2)\psi_{J\sigma}(t_2)
  \end{aligned}
\end{equation}
owhere $\psi_{I\sigma}(t) =\psi_{i_1\sigma}(t) \cdots \psi_{i_{\frac{q}{2}\sigma}}(t)$, note that if $\frac{q}{2}$ is an odd number we have $\psi_{I}$ fermionic that gives us a prefactor $(-1)^{\frac{q}{2}}$ after commuting. 
\begin{equation}
\begin{aligned}
  \ii &S_{\mathrm{int},v,\sigma} = \\
  &\frac{\ii^{2} (-1)^{\frac{q}{2}}}{N^{q-1}} v^2 \frac{2}{q}\int_{t_1 t_2} \sum_{I J} \psi_{\bar{I}\sigma}(t_1)\overline{\psi}_{I\sigma}(t_2)\psi_{J\sigma}(t_2)\overline{\psi}_{\bar{J}\sigma}(t_1)
\end{aligned}
\end{equation}
Define bilocal field $ \ii G_{\sigma}(t_1,t_2) = \frac{1}{N}\sum_{i} \psi_{i\sigma}(t_1)\overline{\psi}_{i\sigma}(t_2)$. Note that rearranging $\psi_{\bar{I}}\psi_{I}$ to pairs will not give additional signs. Here we further assume spin space diagonal and symmetric solution $G_{\uparrow}=G_{\downarrow}$, which allows us to drop the spin index, and divide the bosonic part by $2$ to get per-spin effective action. The $v$ part now becomes
\begin{equation}
   \frac{1}{N} \ii S_{\mathrm{int},v} =   \int_{t_1 t_2}v^2 \frac{2}{q}[\ii G(t_1,t_2)]^{\frac{q}{2}} [-\ii G(t_2,t_1)]^{\frac{q}{2}} 
\end{equation}
which agrees with previous studies~\cite{Gu2019} up to imaginary unit $i$ that from definitions.
The boson-fermion part $\ii S_{\mathrm{int},g}$ is along the same line and straightforward. Restore the time dependence of the couplings. The effective action per spin is then
\begin{equation}
  \begin{aligned}
     \frac{1}{N } &\ii S_{\mathrm{eff}} =\\
     & \tr \log (G_{0}^{-1}-\Sigma) - \frac{\lambda}{2} \tr \log(D_0^{-1}-\Pi)                         \\
     & +  \int_{t_1,t_2} v(t_1)v(t_2) \frac{2}{q} \left[G(t_1,t_2)\right]^{\frac{q}{2}} [G(t_2,t_1)]^{\frac{q}{2}}               \\
     & + \ii \int_{t_1,t_2} g(t_1)g(t_2) \frac{2}{q}  \left[G(t_1,t_2)\right]^{\frac{q}{2}} [G(t_2,t_1)]^{\frac{q}{2}} D(t_1,t_2) \\
     & +  \int_{t_1,t_2} \Sigma(t_1,t_2) G(t_2,t_1)
    - \frac{\lambda}{2} \int_{t_1,t_2} \Pi(t_2,t_1) D(t_1,t_2)
  \end{aligned}
\end{equation}
The saddle point equations are
\begin{equation}
  \label{eq:saddle_eq_0}
  \begin{aligned}
    G(t_1,t_2) =      & \left[G_{0}^{-1}-\Sigma\right]^{-1}(t_1,t_2)                                                                                             \\
    \Sigma(t_1,t_2) = & \left(v(t_1)v(t_2)+\ii g(t_1)g(t_2) D(t_1,t_2) \right)\\ 
    & \times \left[G(t_1,t_2)\right]^{\frac{q}{2}} [G(t_2,t_1)]^{\frac{q}{2}-1}  \\
    D(t_1,t_2) =      & \left[D_{0}^{-1}-\Pi\right]^{-1}(t_1,t_2)                                                                                                \\
    \Pi(t_1,t_2) =    &  \frac{-2\ii}{\lambda}  \frac{2}{q} g(t_1) g(t_2)\left[G(t_1,t_2)\right]^{\frac{q}{2}} \left[G(t_2,t_1) \right]^{\frac{q}{2}}
  \end{aligned}
\end{equation}
For general $q$ we may define the $q$ dependent couplings which should be finite when $q\to \infty$. This can be reached by $g^2 = \mathcal{J}^2 \frac{2^{q-1}}{q}$ and $\lambda = \frac{4}{q^2} \lambda_1$ with fixed $\mathcal{J}$ and $\lambda_1$. We then have the boson coupling $\frac{g^2}{\lambda}\frac{2}{q} = \mathcal{J}^2 \frac{2^{q}}{q^2} \frac{q^2}{4\lambda_1} = \mathcal{J}^2 2^{q-2}$ which is uniform in $q$ as well. The lattice coupling gives a complex SYK$_q$ model. For given $q$, the full model is a SYK$_q$ plus YSYK$_q$ model. Note that due to the bosonic propagator, the YSYK$_2$ is closely related to SYK$_4$, which means for $q=2$ it may share some features of SYK$_2$-SYK$_4$ models.

\section{More on equilibrium properties}
\label{eq:more_on_equilibrium_properties}
% \begin{table}
%   \caption{Scaling exponent $\Delta$ for different $q$ with $\lambda=1$.}
%   \label{tab:exponent_table2}
%   \begin{tabular}{|c|c|c|c|c|c|c|}
%   \hline
%   $q$      & 2        & 4        & 6        & 8        & 10             \\ \hline
%   $\Delta$ & 0.420374 & 0.365563 & 0.332526 & 0.313157 & 0.30086  \\ \hline
%   \end{tabular}
%   \end{table}
%   % {0.420374, 0.365563, 0.332526, 0.313157, 0.30086}

\begin{figure}
  \centering
  \includegraphics[width=0.9\linewidth]{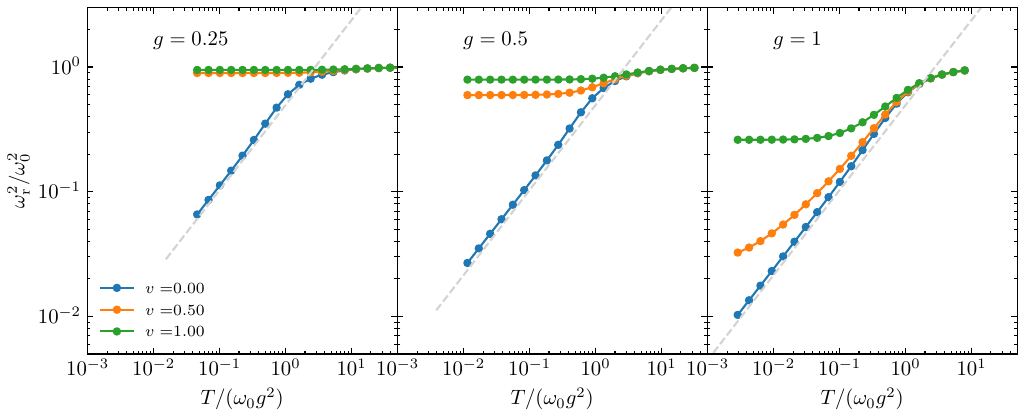}
  \caption{Equilibrium boson squared renormalized frequency $\omega_{\mathrm{r}}^2$ vs temperature $T$, $q=2$. The dashed lines are $\sim T^{4\Delta-1}$, i.e., the expected low-temperature behavior. Here, $10^4$ Matsubara frequencies are used for the numerical solutions mainly because of those low target temperatures.}
  \label{fig:eq_wr}
\end{figure}

In Fig.~\ref{fig:eq_wr} we numerically compute the renormalized boson frequency, defined as $\omega_{\mathrm{r}}^2 = \omega_0^2 + \re \Pi(\omega_n=0)$ with iterative method as that is used in Fig.~\ref{fig:eq_G_D_vs_conformal}. This actually reproduces the known result in the literature~\cite{Valentinis2023}, where the $v=0$ curve is following the universal scaling $\omega_{\mathrm{r}}^2 \sim T^{4\Delta-1}$. As we have also seen from Fig.~\ref{fig:phase_diagram_q2}, for $q=2$ the effect of increasing the finite $v$ is to introduce a SYK$_2$ Fermi liquid phase that covers an increasing $g$ interval and pushes the non-Fermi liquid phase to the right, i.e., the larger $g$. This can be seen from Fig.~\ref{fig:eq_wr} that for $g=1$ and $v=0.5$ the renormalized phonon frequencies are softened considerably by lowering the temperature until the SYK$_2$ Fermi liquid at very low temperature. This effect can be used to tune the phase from NFL to FL without changing the temperature, and this may be done dynamically through quantum quenches.

\section{Quench with fixed lattice coupling $v$}
\label{sec:fix_v_quench}

Here we consider the Yukawa interaction to quench
\begin{equation}
v \to v,  \quad g_i \to g_f
\end{equation}
which means we fix the lattice coupling $v$ and quench the YSYK coupling to a different value after the quench time point $t_0=0$. The quenches are applied through a theta function $\theta(t_0)$ which enters the $g(t_1)g(t_2)$ terms of the saddle point equations and thus the Kadanoff-Baym equations.

\subsection{Spectral functions and distribution}
\begin{figure}
  \centering
  \begin{overpic}[width=0.425\linewidth]{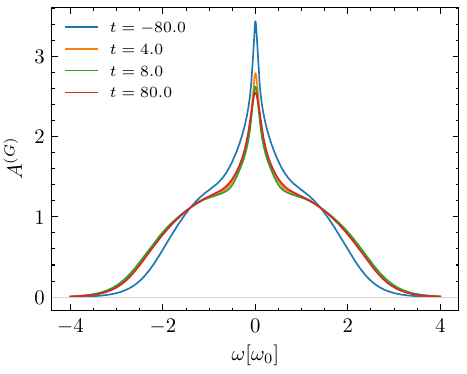}
    \put(90,70){a} 
  \end{overpic}
  \begin{overpic}[width=0.45\linewidth]{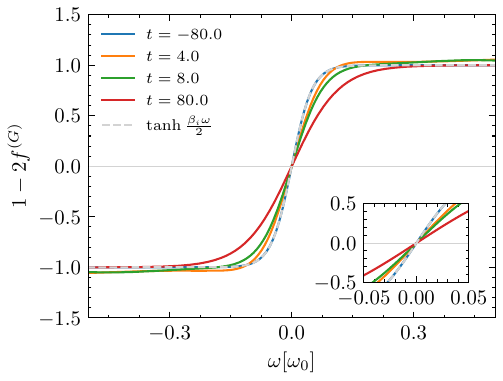}
    \put(90,65){b} 
  \end{overpic}
  \begin{overpic}[width=0.45\linewidth]{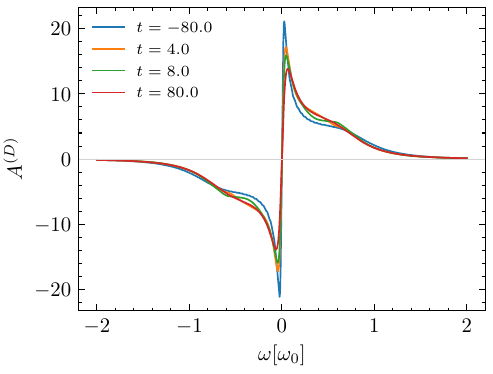}
    \put(90,65){c} 
  \end{overpic}
  \begin{overpic}[width=0.45\linewidth]{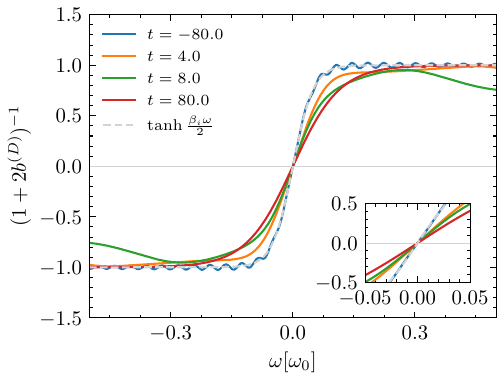}
    \put(90,65){d}
  \end{overpic}
  \caption{Spectral functions and effective distributions for fixed. Fixed $v$ quench where $v=0$, $g_i=1$, $g_f=1.1$ and $q=2$. The insets of (c) and (d) are zoomed-in pictures. Here we can see a sudden width change of the spectral function, which is because the $g$, which is controlling the width, is quenched.}
  \label{fig:fixv_q2_spectral}
\end{figure}

In Fig.~\ref{fig:fixv_q2_spectral} we show the time dependent spectral function $A^{(G)}(\omega,t_{\mathrm{a}})$, $A^{(D)}(\omega,t_{\mathrm{a}})$ and effective distributions $f^{(G)}(\omega,t_{\mathrm{a}})$ and $b^{(D)}(\omega,t_{\mathrm{a}})$ for fixed $v=0$ and a sudden quench of $g$ from $1.0$ to $1.1$. Note that the superscripts on the observables indicate which two-point function they are extracted from. From the plots, one can see that the fermion spectral function has a jump on its bandwidth, which is because of the sudden quench of the $g$, which determines the bandwidth especially when $v=0$. From the $A^{(G)}$ and $ A^{(G)}$ we can see that the peak heights of the spectral functions decreased and converged to the final value quickly after the quench, and the overall shape remains almost unchanged which suggests a quasi-equilibrium process since it is known that the low-frequency peak heights are reflecting the temperatures in equilibrium. Similar behaviors can be found for the effective distributions $f^{(G)}$ and $b^{(D)}$ plots, where the slope around $\omega=0$ can be regarded as an effective temperature. For instance, at $t=8$, both the spectral functions and distributions of the fermions and bosons are close to the final value, but still have obvious deviations. This suggests that the boson and fermion observables thermalize on a comparable timescale. 

\subsection{Dynamics of boson occupations, renormalized frequencies, and correlation energies}

\begin{figure}
  \centering
  \begin{overpic}[width=0.43\linewidth]{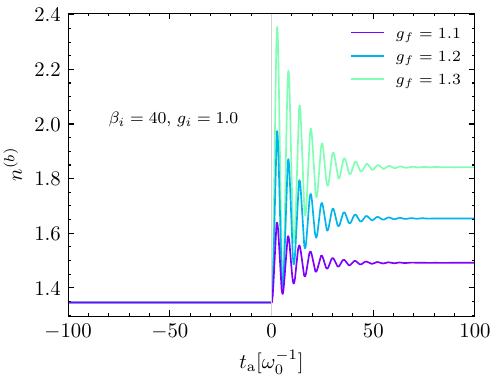}
    \put(18,65){a}
  \end{overpic}
  \begin{overpic}[width=0.45\linewidth]{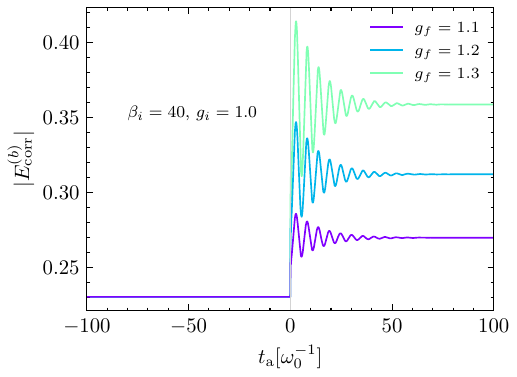}
    \put(20,62){b}
  \end{overpic}
  \begin{overpic}[width=0.45\linewidth]{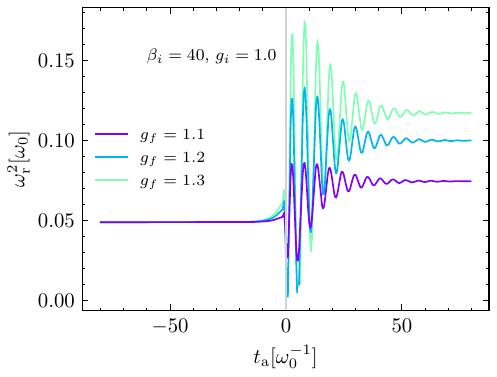}
    \put(20,67){c}
  \end{overpic}
  \begin{overpic}[width=0.44\linewidth]{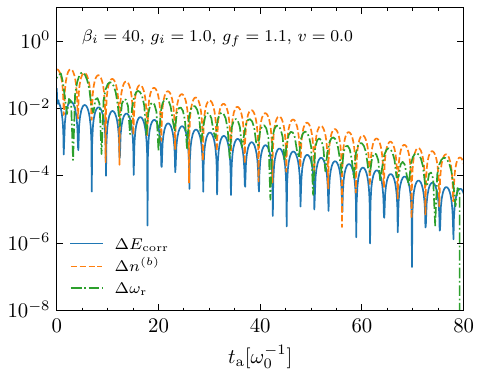}
    \put(85,67){d}
  \end{overpic}
  \caption{From (a-c), boson occupations, correlation energies, renormalized frequencies. Fixed $v=0$, $q=2$. Note that $\omega_{\mathrm{r}}^2$ is extracted from Fourier space using the Wigner coordinate two-point function, thus, some precursors occur. In (d), we compare the quantities of (a-c) with $g_f=1.1$ to see if their time dependence, oscillation frequencies, and decay rates are very similar to each other.
  } 
  \label{fig:fixv_q2_nbEcorrwr}
\end{figure}

In Fig.~\ref{fig:fixv_q2_nbEcorrwr} we show boson occupations $n^{(b)}(t_{\mathrm{a}})$, boson renormalized frequencies $\omega_{\mathrm{r}}(t_{\mathrm{a}})$ and boson correlation energies $E_{\mathrm{corr}}^{(b)}(t_{\mathrm{a}})$ for the fixed $v$ quenches. The quench dynamics of those quantities gives under-damped oscillations.

\begin{figure}
  \centering
  \begin{overpic}[width=0.95\linewidth]{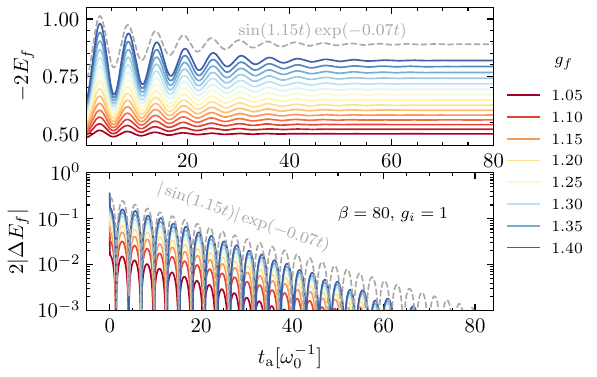}
    \put(78,57){a}
    \put(78,30){b}
  \end{overpic}
  \caption{The real time correlation energies $E^{(f)}_{\mathrm{corr}}(t_{\mathrm{a}})$ (in plots labeled as $E_f$) from fermion propagators and self-energies through Galitskii-Migdal formula. We defined $\Delta E^{(f)} = E^{(f)}(t)-E_{f}^{(f)}$. The early to intermediate time decay rate is about $0.07\omega_0$ and almost the same for different final coupling strengths. The oscillation frequencies are also insensitive to different $g_f$ and are about $1.15\omega_0$. Here $v=0$ and $q=2$. Note here the quench is applied immediately at the initial time $t_0=t_i$ and we used $N_{\tau}=600$ and $h=0.02$, with initially $\beta=80$.}
  \label{fig:energy_q_2}
\end{figure}

In Fig.~\ref{fig:energy_q_2} we show the correlation energy extracted from the fermion two-point functions. Use the observation that the kinetic energy purely from boson is where $E_{\mathrm{kin}}(t) \sim \partial_t^2 \braket{\phi(t)\phi(t)}$, which then shares the same frequency with the boson occupation oscillations. Since the total energy is conserved after the quench $E_{\mathrm{kin}}(t)+E_{\mathrm{corr}}(t)=E_{\mathrm{total}}$, the correlation energy should oscillates with the frequency $\omega_{\phi^2}$ as well. The oscillations may be understood from the discussions in Sec.~\ref{sec:occupation_equal_time_equations} in the main text.

\subsection{Time dependent effective temperature}
\begin{figure}
  \centering
  \begin{overpic}[width=0.45\linewidth]{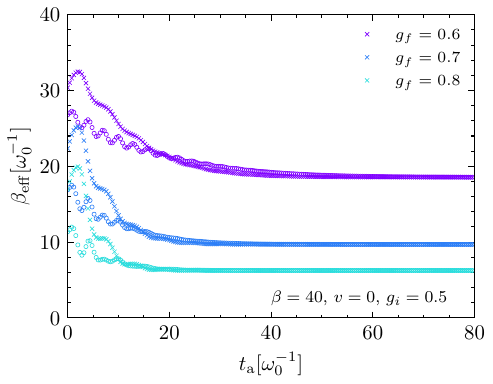}
    \put(1,70){a}
  \end{overpic}
  \begin{overpic}[width=0.45\linewidth]{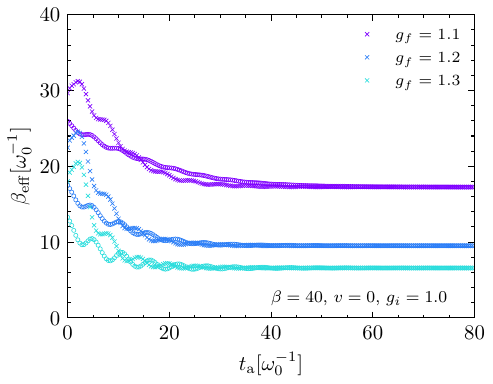}
    \put(1,70){b}
  \end{overpic}
  \begin{overpic}[width=0.45\linewidth]{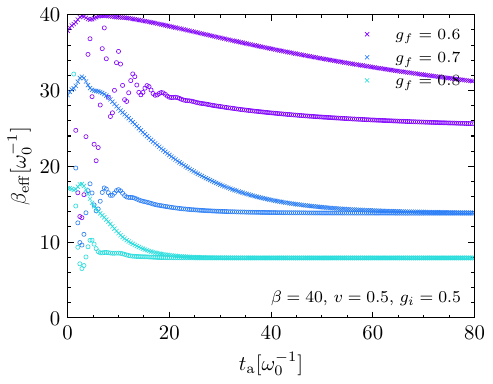}
    \put(1,70){c}
  \end{overpic}
  \begin{overpic}[width=0.45\linewidth]{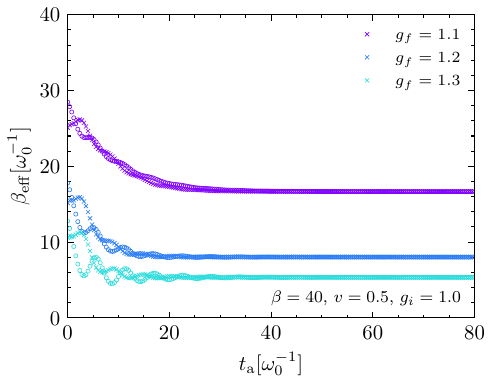}
    \put(1,70){d}
  \end{overpic}
  \begin{overpic}[width=0.45\linewidth]{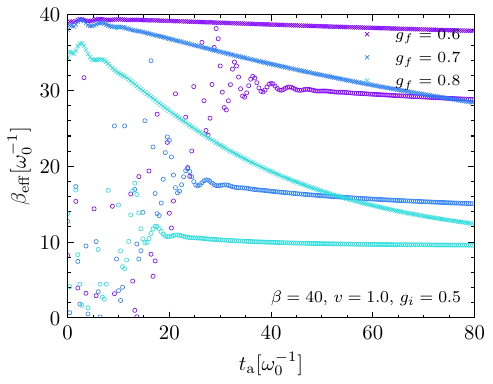}
    \put(1,70){e}
  \end{overpic}
  \begin{overpic}[width=0.45\linewidth]{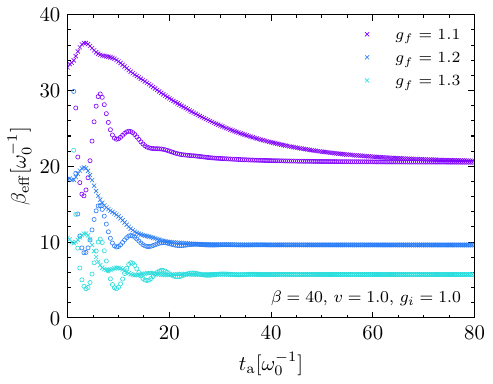}
    \put(1,70){f}
  \end{overpic}
  \caption{Time dependence of effective inverse temperatures for $q=2$. Here, crosses are for fermions and circles for bosons. Left column (a, c, e) $g=0.5$ and right column (b, d, f) $g=1.0$. The rows are $v=0,0.5,1.0$ respectively. From (a, b, d, f) we can see similar temperature dynamics as the $v$ quench. From (c) and (e), we can see that we cannot reach the equilibrium time scale because the thermalization between the boson and fermion subsystems is much slower when quenched from the Fermi liquid phase. The distribution functions cannot be described by a single parameter for the latest time we can reach. This can be regarded as prethermalization. Note that the strong oscillations for short times indicate the effective temperatures may not be well-defined.}
  \label{fig:fixv_q2_beta}
\end{figure}

In Fig.~\ref{fig:fixv_q2_beta} we plot the time-dependent effective inverse temperatures $\beta^{(G)}(t_{\mathrm{a}})$ and $\beta^{(D)}(t_{\mathrm{a}})$ extracted from boson and fermion propagators through a symmetric numerical differentiation around $\omega=0$ of $1-2f^{(G)}(\omega,t_{\mathrm{a}})$ and $(1+2b^{(D)}(\omega,t_{\mathrm{a}}))^{-1}$. The left column (a, c, e), shows $g_i=0.5$, and the right column (b, d, f), shows $g_i=1.0$ results. The rows have constant $v=0,0.5,1.0$ respectively. 

From all plots, we can see that the quench affects two subsystems, the bosons and the fermions, differently. For $v=0$, Fig.~\ref{fig:fixv_q2_beta} (a, b) the fermion inverse temperature $\beta^{(G)}(t_{\mathrm{a}})$ decays slower than the boson, however, the initial post-quench inverse temperature of the fermion is larger than the boson, which means the fermion temperature is initially lower than boson temperature. There are also oscillations in the fermion and boson inverse temperature. However, with different periodicity, those for fermions oscillate more slowly than bosons. For $v=0.5$, in (c) we can see that since the starting state is closer to the Fermi liquid phase, the equilibration time when the two temperatures equal is later. For $g_f=0.6$, the effective temperatures of fermions and bosons are not equal to each other even for the latest computed time $t_{\mathrm{a}}=80$. Also, there are strong oscillations for bosons that signal temperatures are not well defined. For $v=1.0$, Fig.~\ref{fig:fixv_q2_beta} (e, f), we can see that for $g_i=0.5$ and $g_f=0.6$, where the quench is deep inside the Fermi liquid phase, the late time $\beta_{\mathrm{eff}}(t_{\mathrm{a}})$ for bosons and fermions are almost parallel.

\begin{figure}
  \centering
  \begin{overpic}[width=0.45\linewidth]{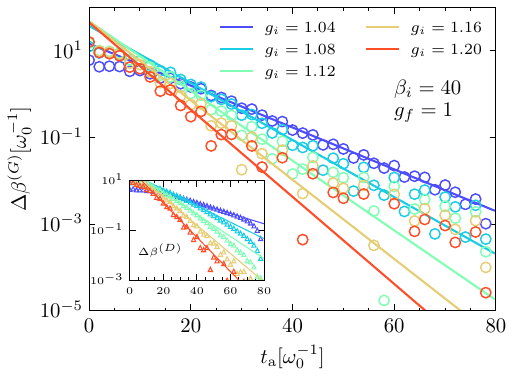}
  \put(1,65){a}
  \end{overpic}
  \begin{overpic}[width=0.45\linewidth]{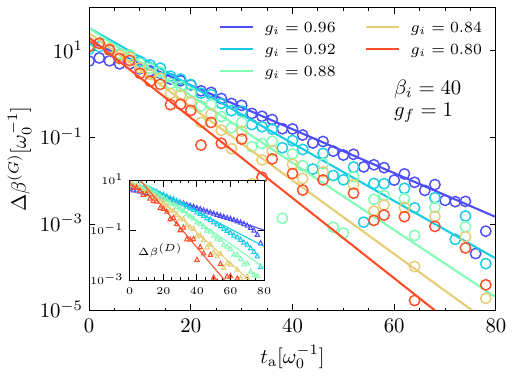}
  \put(1,65){b}
  \end{overpic}
  \begin{overpic}[width=0.45\linewidth]{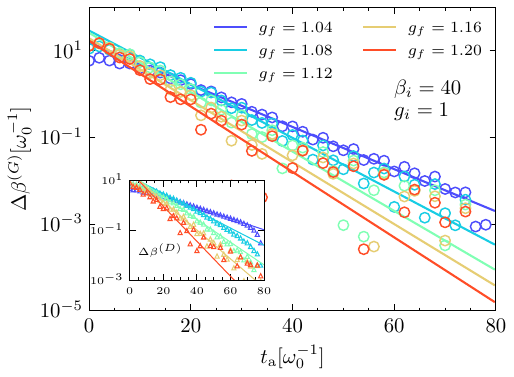}
  \put(1,65){c}
  \end{overpic}
  \begin{overpic}[width=0.45\linewidth]{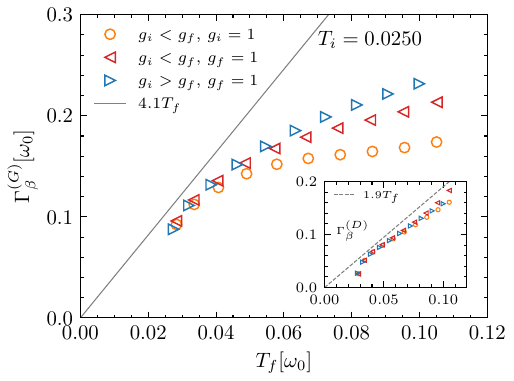}
  \put(1,65){d}
  \end{overpic}
  \caption{Temperature relaxations. Fixed $v=0$, $q=2$. $\Delta \beta(t) = \beta(t)-\beta_f$. Here, markers are data, and solid lines are fitted curves or guides of the eye. Here (a) $g_i> g_f$, (b) $g_i<g_f$ and $g_f=1$, (c) $g_i<g_f$ and $g_i=1$. The relaxation rates for (a, b, c) are summarized in (d), and the inset of (d) shows the results for bosons. One can see that boson relaxations are linear in $T_f$ and close to each other for the three protocols, and also close to the $v$ quench slope $1.9T_f$. Fermions show different dynamics.
  }
  \label{fig:fixv_q2_v0_Gamma}
\end{figure}

In Fig.~\ref{fig:fixv_q2_v0_Gamma}, the inverse temperature relaxation rates $\Gamma_{\beta}$ are shown, where they are computed by using the same technique as that for Fig.~\ref{fig:fixg_q2_Gamma}. The initial temperature is set to $\beta=40$. Three different protocols are considered here. They are shown in Fig.~\ref{fig:fixv_q2_v0_Gamma}. In (a), we show the fits for the quench from larger Yukawa coupling to smaller ones $g_i> g_f$. In (b) $g_i<g_f$ and $g_f=1$. In (c) $g_i<g_f$ and $g_i=1$. Those protocols are for studying the effect of being quenched from different sides of $g=1$ and to see if they are universal. From (a) to (c), we find that there are stronger oscillations for the fermion effective inverse temperatures, which make the fit harder than the $v$ quenches. In (d), we summarize the relaxation rates. One can see the difference compared with the $v$ quenches, especially from the rates of fermions. 

\subsection{Quantum quenches for $q\geq 4$ with $M=N/q^2$}

We use the scaling
\begin{equation}
  g_{q|i,f} = g\sqrt{\frac{2^{q-1}}{q}},  \quad v_{q|i,f} = v\sqrt{\frac{2^{q-1}}{q}},\quad \lambda_{q} = \lambda \frac{4}{q^2}
\end{equation}
to get a more uniform $q$ dependence.

\begin{figure}
  \centering
  \begin{overpic}[width=0.45\linewidth]{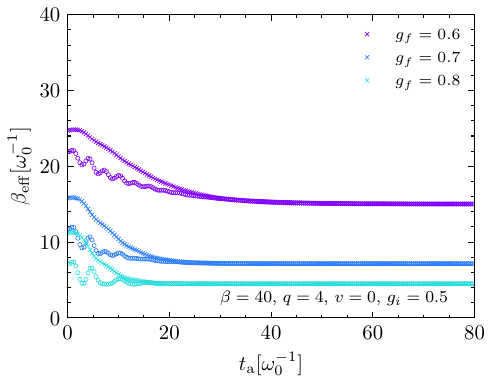}
    \put(1,69){a}
  \end{overpic}
  \begin{overpic}[width=0.45\linewidth]{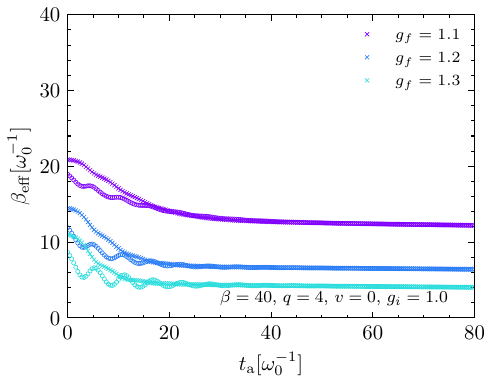}
    \put(1,69){b}
  \end{overpic}
  \begin{overpic}[width=0.45\linewidth]{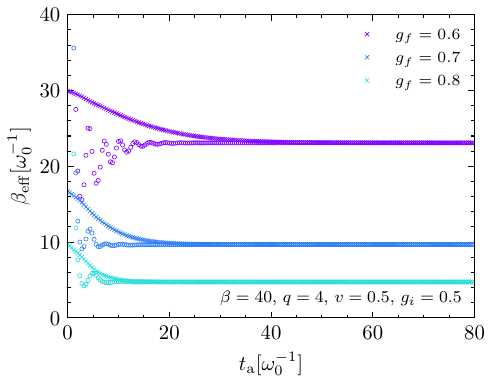}
    \put(1,69){c}
  \end{overpic}
  \begin{overpic}[width=0.45\linewidth]{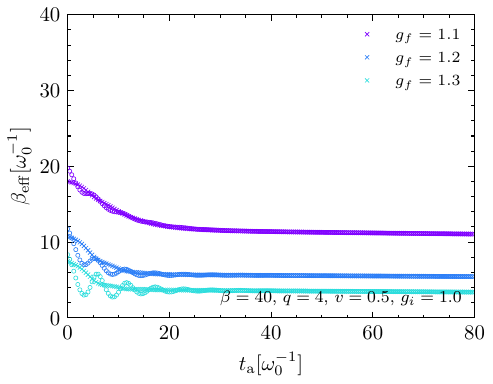}
    \put(1,69){d}
  \end{overpic}
  \begin{overpic}[width=0.45\linewidth]{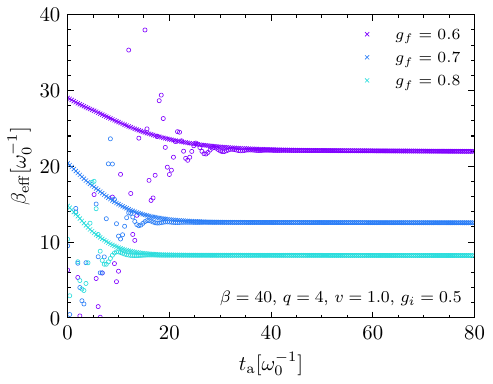}
    \put(1,69){e}
  \end{overpic}
  \begin{overpic}[width=0.45\linewidth]{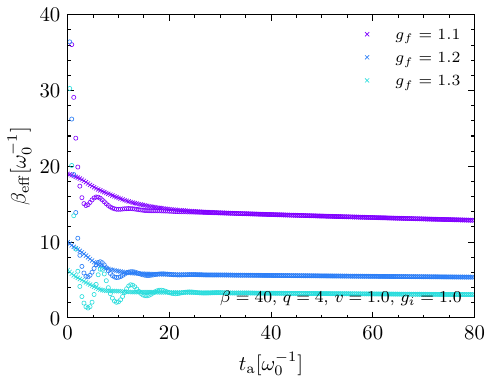}
    \put(1,69){f}
  \end{overpic}
  \caption{Time dependence of effective inverse temperatures for $q=4$. Here, solid lines are for fermions and circles for bosons. Left column $g=0.5$ and right column $g=1.0$. The rows are $v=0,0.5,1.0$ respectively. Compare (c) and (e) with the $q=2$ results in Fig.~\ref{fig:fixv_q2_beta} (c) and (e), we can see that for $q=4$ the temperature relaxations are much faster, which is because of the $q=4$ SYK lattice term.}
  \label{fig:fixv_q4_beta}
\end{figure}

In Fig.~\ref{fig:fixv_q4_beta} we present the results for $q=4$ where the lattice part is an SYK$_4$. Otherwise, the parameters are similar to those in Fig.~\ref{fig:fixv_q2_beta}. If we look at the (c) and (e), we can find that the fermions thermalize much faster in the $q=4$ case, which is expected because the $v$ terms are SYK$_{4}$.

\section{Details about the kinetic equations for scale separation}
\label{sec:kinetic_eq_details}
\subsection{Approximate kinetic equations}

Use the Keldysh component of the Dyson equation
\begin{equation}
  \left[G^{-1}\right]^R \circ G^K \circ\left[G^{-1}\right]^A=\Sigma^K 
\end{equation}
and apply the parametrization $G^K=G^R \circ F-F \circ G^A$ we get
\begin{equation}
 F \circ G_0^{-1}-G_0^{-1} \circ F=\Sigma^K- (\Sigma^R \circ F-F \circ \Sigma^{A})
  \end{equation}
where we dropped the $R, A$ of the inverse free Green's function since they are different only by a $\pm \ii 0^{+}$ for causality, which won't affect the equations here. Use $[A\overset{\circ}{,} B] = A\circ B - B\circ A$ it is 
\begin{equation}
  -[G_0^{-1}\overset{\circ}{,} F]=\Sigma^K- (\Sigma^R \circ F-F \circ \Sigma^{A})
\end{equation}
The boson equation is the same 
\begin{equation}
  -[D_0^{-1}\overset{\circ}{,} B]=\Pi^K- (\Pi^R \circ B-B \circ \Pi^{A})
\end{equation}
% Kamenev2011 (5.24), section 5.7 boson
% Kamenev2011 (9.67)
Use the Wigner transform identities~\cite{Kamenev2011}
\begin{equation}
[A \overset{\circ}{,} B] \xrightarrow{\mathrm{WT}} \ii (\partial_{t}A \partial_{\omega}B-\partial_{\omega}A \partial_{t}B)+\dots
\end{equation}
\begin{equation}
A \circ B \xrightarrow{\mathrm{WT}}  A B + \dots
\end{equation}
which gives
\begin{equation}
\left[(\mathrm{i} \partial_t+\mu)\overset{\circ}{,} F\right] \xrightarrow{\mathrm{WT}} \mathrm{i} \partial_\omega \omega \partial_t F=\mathrm{i} \partial_t F
\end{equation}
\begin{equation}
\left[(-\partial_t^2-\omega^2) \overset{\circ}{,} B\right] \xrightarrow{\mathrm{WT}} \mathrm{i} \partial_\omega \omega^2 \partial_t F=2 \mathrm{i} \omega \partial_t F 
\end{equation}
and thus the equations in Wigner transformed space 
\begin{equation}
  -\ii \partial_t F=\Sigma^K- F(\Sigma^R  -  \Sigma^{A})
\end{equation}
\begin{equation}
  -2\ii \omega \partial_t B=\Pi^K- B(\Pi^R - \Pi^{A})
\end{equation}
Note the functions are with variables $(\omega,t_{\mathrm{a}})$.

\subsection{Details about collision integrals}
The collision integrals can be written as
\begin{equation}
  I_{\mathrm{col}}^{(f)} = \Sigma^{K}(\omega,t_{\mathrm{a}})-F(\omega,t_{\mathrm{a}})(\Sigma^>(\omega,t_{\mathrm{a}})-\Sigma^<(\omega,t_{\mathrm{a}}))
\end{equation}
\begin{equation}
  I_{\mathrm{col}}^{(b)} = \Pi^{K}(\omega,t_{\mathrm{a}}) - B(\omega,t_{\mathrm{a}})(\Pi^>(\omega,t_{\mathrm{a}})-\Pi^<(\omega,t_{\mathrm{a}}))
\end{equation}
We can use the exact relations $
  \Sigma^{K}(\omega,t_{\mathrm{a}}) = \Sigma^>(\omega,t_{\mathrm{a}})+\Sigma^<(\omega,t_{\mathrm{a}})
$ and
$
  \Pi^{K}(\omega,t_{\mathrm{a}}) = \Pi^>(\omega,t_{\mathrm{a}})+\Pi^<(\omega,t_{\mathrm{a}})
$ then only greater and lesser functions are required. Under the scale separation assumption, we can assume Green's functions have the form
\begin{equation}
  G^>(\omega,t_{\mathrm{a}}) = -\ii A^{(G)}\frac{F+1}{2} =  -\ii A^{(G)}(1-f)
\end{equation}
\begin{equation}
  G^<(\omega,t_{\mathrm{a}}) = -\ii A^{(G)}\frac{F-1}{2} = \ii A^{(G)} f
\end{equation}
\begin{equation}
  D^>(\omega,t_{\mathrm{a}}) = -\ii A^{(D)}\frac{B+1}{2} = -\ii A^{(D)}(1+b)
\end{equation}
\begin{equation}
  D^<(\omega,t_{\mathrm{a}}) = -\ii A^{(D)} \frac{B-1}{2} = -\ii A^{(D)} b
\end{equation}
where $F= 1-2f$ and $B=1+2b$.

We then recall that the self-energies are
\begin{equation}
  \begin{aligned}
    \Sigma(t_1,t_2) & =  R_{q-2}(t_1,t_2) (v^2 + g^2\ii D(t_1,t_2))   G(t_1,t_2)                 \\
    \Pi(t_1,t_2)    & =  \frac{-4\ii g^2}{\lambda q} R_{q-2}(t_1,t_2) G(t_1,t_2) G(t_2,t_1)
  \end{aligned}
\end{equation}
with
$
  R_{q-2}(t_1,t_2) \equiv [G(t_1,t_2)G(t_2,t_1)]^{\frac{q-2}{2}}
$. Since we are considering the quenches, we suppress the time dependence of the parameters. They can be restored whenever it is necessary.
We define the frequency convolution
\begin{equation}
  [A \bullet  B](\omega) = \int_{\omega} A(\nu-\omega) B(\omega)
\end{equation}
where $\int_{\omega} = \int \dd \omega/2\pi$. For the terms inside the self-energies, we have
\begin{equation}
  A(t_1,t_2) B(t_1,t_2) \xrightarrow{\mathrm{WT}}  A \bullet B
\end{equation}
One can then write down the Wigner-transformed self-energies
\begin{equation}
  \Sigma^{>,<}(\omega,t_{\mathrm{a}}) = R_{q-2}^{>,<} \bullet (v^2 + g^2 \ii D^{>,<})\bullet  G^{>,<}      
\end{equation}
\begin{equation}
  \Pi^{>,<}(\omega,t_{\mathrm{a}}) = \frac{-4 \ii g^2}{\lambda q} R_{q-2}^{>,<} \bullet G^{>,<} \bullet G^{<,>}_{-}
\end{equation}
\begin{equation}
  R_{q-2}^{>,<}(\omega,t_{\mathrm{a}}) =\bullet^{\frac{q-2}{2}} [G^{>,<}\bullet G^{<,>}_{-}]
\end{equation}
where $\bullet^{\frac{q-2}{2}}[A] = A \bullet A \bullet \cdots \bullet A $ with total $\frac{q-2}{2}$ terms.
where $G_{-}$ is the Wigner half-fourier transformed function $G_{-}(\omega,t_{\mathrm{a}})=[G(t_2,t_1)](\omega,t_{\mathrm{a}})$. 

We rewrite the Green's functions in terms of the spectral functions and distribution functions
\begin{equation}
  \begin{aligned}
    &\Sigma^{>}(\omega,t_{\mathrm{a}})
    = R_{q-2}^{>} \bullet(v^2 + g^2 \ii D^{>}) \bullet G^>\\ 
    &= R_{q-2}^{>} \bullet \left(v^2 + g^2 A^{(D)}\frac{B+1}{2} \right)\bullet \left(-\ii A^{(G)}\frac{F+1}{2} \right)      \\
  \end{aligned}
  \end{equation}
\begin{equation}
  \begin{aligned}
    &\Sigma^{<}(\omega,t_{\mathrm{a}}) = R_{q-2}^{<} \bullet (v^2 + g^2 \ii D^{<})\bullet  G^{<}    \\  
    &=R_{q-2}^{<} \bullet \left(v^2 + g^2 A^{(D)} \frac{B-1}{2}\right) \bullet  \left(-\ii A^{(G)}\frac{F-1}{2} \right)
  \end{aligned}
  \end{equation}
\begin{equation}
  \begin{aligned}
    &\Pi^{>}(\omega,t_{\mathrm{a}}) = \frac{-4 \ii g^2}{\lambda q} R_{q-2}^{>} \bullet G^{>} \bullet G^{<}_{-}\\
    &=\frac{-4 \ii g^2}{\lambda q} R_{q-2}^{>} \bullet \left( A^{(G)}\frac{F+1}{2}\right) \bullet \left( A^{(G)}_{-}\frac{1-F_{-}}{2}\right)
  \end{aligned}
  \end{equation}
\begin{equation}
  \begin{aligned}
    &\Pi^{<}(\omega,t_{\mathrm{a}}) = \frac{-4 \ii g^2}{\lambda q} R_{q-2}^{<} \bullet G^{<} \bullet G^{>}_{-}\\
    &=\frac{-4 \ii g^2}{\lambda q} R_{q-2}^{<} \bullet \left( A^{(G)}\frac{1-F}{2}\right) \bullet \left( A^{(G)}_{-}\frac{F_{-}+1}{2}\right)
  \end{aligned}
\end{equation}
\begin{equation}
  \begin{aligned}
    &R_{q-2}^{>}(\omega,t_{\mathrm{a}}) =\bullet^{\frac{q-2}{2}} [G^{>}\bullet G^{<}_{-}]\\
    &=\bullet^{\frac{q-2}{2}} \left[\left( A^{(G)}\frac{F+1}{2}\right) \bullet \left( A^{(G)}_{-}\frac{1-F_{-}}{2}\right) \right]
  \end{aligned}
\end{equation}
\begin{equation}
  \begin{aligned}
    &R_{q-2}^{<}(\omega,t_{\mathrm{a}}) =\bullet^{\frac{q-2}{2}} [G^{<}\bullet G^{>}_{-}]\\
    &=\bullet^{\frac{q-2}{2}} \left[  \left(A^{(G)}\frac{1-F}{2}\right)\bullet\left( A^{(G)}_{-}\frac{F_{-}+1}{2}\right)\right]
  \end{aligned}
\end{equation}
Note $A_{-}= A(-\omega,t_{\mathrm{a}})$.

\subsection{$q=2$}

We first consider $q=2$, which means we can set $R_{q-2}=1$. The results are listed in \eqref{eq:Ifcoll_q_equal_2} and \eqref{eq:Ibcoll_q_equal_2}
\begin{widetext}
  \begin{equation}
    \label{eq:Ifcoll_q_equal_2}
    I_{\mathrm{col}}^{(f)} = \frac{\ii g^2}{2} \left(
   F \left( (A^{(D)} B) \bullet  A^{(G)}+A^{(D)} \bullet ( A^{(G)} F) \right)
   - (A^{(D)} B) \bullet ( A^{(G)} F)
   - A^{(D)} \bullet  A^{(G)}
  \right)
\end{equation}
\begin{equation}
  \label{eq:Ibcoll_q_equal_2}
  I_{\mathrm{col}}^{(b)} = 
  \frac{\ii g^2}{\lambda}\left(
  F \left( ( A^{(G)} F) \bullet  A^{(G)}_{-} -A^{(G)} \bullet ( A^{(G)}_{-} F_{-} ) \right)
  + ( A^{(G)} F) \bullet ( A^{(G)}_{-} F_{-} )
  -  A^{(G)} \bullet A^{(G)}_{-} 
  \right)
\end{equation}
\end{widetext}
One can see that there is no $v^2$ term for the $q=2$ collision integral, similar to that in~\cite{Larzul2022}. This can be explained by that the SYK$_2$ is a free fermion model.

\subsection{$q>2$ collision integrals}

Unlike that of $q=2$, for $q>2$ the $v^2$ terms contribute to the collision integral. The collision integrals for $q>2$ are lengthier while still straightforward, we list them here in the equations \eqref{eq:Ifcoll_q_greater_2} and \eqref{eq:Ibcoll_q_greater_2}. The main difference is that the "rung" function $R$ contributes to the collision integral, where the pure SYK term, the $v^2$ terms, become collisional. This suggests that the relaxation dynamics of $q=2$ and $q>2$ should be very different.
\begin{widetext}
  \small
  \begin{equation}
    \label{eq:Ifcoll_q_greater_2}
    I^{(f)}_{\mathrm{coll}}=
      \frac{1}{2} \ii (F-1)  R^>_{q-2}  \bullet \left(\frac{1}{2}  A^{(D)} (B+1) g^2+v^2\right) \bullet ( A^{(G)} (F+1))
    -  \frac{1}{2} \ii (F+1)  R^<_{q-2}  \bullet \left(\frac{1}{2}  A^{(D)} (B-1) g^2+v^2\right) \bullet ( A^{(G)} (F-1))
  \end{equation}
  \begin{equation}
    \label{eq:Ibcoll_q_greater_2}
    I^{(b)}_{\mathrm{coll}}=
    \frac{\ii g^2}{\lambda  q}\left( ((F-1)  R^>_{q-2}  \bullet ( A^{(G)} F+ A^{(G)}) \bullet ( A^{(G)}_{-}- A^{(G)}_{-}  F_{-})
    -(F+1)  R^<_{q-2}  \bullet ( A^{(G)}- A^{(G)} F) \bullet ( A^{(G)}_{-}  F_{-}+ A^{(G)}_{-}))
    \right)
  \end{equation}
\end{widetext}

\subsection{Equations for spectral functions}
Equations for spectral functions can be found by using left and right Dyson equations~\cite{Rammer1986, Larzul2022}
$
  G_{0}^{-1} \circ G^{R,A} = \delta + \Sigma^{R,A} \circ G^{R,A}
$ and 
$
  G^{R,A} \circ G_{0}^{-1}= \delta + G^{R,A}\circ\Sigma^{R,A}  
$ where the $\circ$ is the time-space convolution for two-time functions. The $\delta$ is the delta function for the time difference. The difference between the left and right Dyson equations gives
$
  [G_0^{-1}  \overset{\circ}{,} G^{R} ] = [G^R \overset{\circ}{,} \Sigma^R]
$ and
$
  [G_0^{-1}  \overset{\circ}{,} G^{A} ] = [G^A \overset{\circ}{,} \Sigma^A]
$,
where we used a convolution commutator notation $[A\overset{\circ}{,} B] = A\circ B - B\circ A$. Apply the Wigner transform and gradient expansion to the lowest order we get two equations
\begin{equation}
  -\ii \partial_{t_{\mathrm{a}}} G^{R,A}(\omega,t_{\mathrm{a}}) = 0
\end{equation}
Further difference $R$ and $A$ equations, and take $\mu=0$ gives
\begin{equation}
  \partial_{t_{\mathrm{a}}} A^{(G)}(\omega,t_{\mathrm{a}}) = 0
\end{equation}
Apply the same procedures to the boson we have
\begin{equation}
  -2\ii \omega \partial_{t_{\mathrm{a}}} D^{R,A}(\omega,t_{\mathrm{a}}) = 0
\end{equation}
which means
\begin{equation}
  \partial_{t_{\mathrm{a}}} A^{(D)}(\omega,t_{\mathrm{a}}) = 0 ,\quad \text{If $\omega \neq 0 $}
\end{equation}
For strongly correlated systems, the above equations are not expected to capture the actual spectral function even with the local equilibrium, this is because the spectral functions are sensitive to the distributions. One may assume an instant influence from the distributions like that used in~\cite{Picano2021}, the spectral functions are then determined by consistent solutions
\begin{equation}
  A^{(G,D)}(\omega,t_{\mathrm{a}}) = A^{(G,D)}(\omega,\beta_{\mathrm{eff}}(t_{\mathrm{a}}))
\end{equation}
where the spectral functions are solved from equilibrium self-consistent problem with the input inverse temperatures $\beta_{\mathrm{eff}}(t_{\mathrm{a}})$.

\section{Details about the equal-time equations}
\label{sec:equal_time_eq_details}
Recall the Dyson equation for bosons on the contour
\begin{equation}
  D_0^{-1} \circ_{\mathcal{C}} D = \Pi \circ_{\mathcal{C}} D
\end{equation}
Here $\circ_{\mathcal{C}}$ is the contour convolution $A \circ_{\mathcal{C}}B \equiv \int_{s\in \mathcal{C}} A(t_1,s)B(s,t_2)$.
% D0 has a contour delta 
\begin{equation}
  D_0^{-1}(t_1,t_2) = \left(-\partial_{t_1}^2 -\omega_0^2\right)\delta_{\mathcal{C}}(t_1,t_2)
\end{equation}
where the contour delta function behaves like the usual delta under the contour integral $\int_{x\in \mathcal{C}} \delta_{\mathcal{C}}(x,y) g(x) = g(y)$. The left and right Kadanoff-Baym equations (KBE) on the three-branch contour can be written down by using Langreth rules~\cite{Aoki2014, Stefanucci2013}. Here we consider the KBE for bosons and the adjoint equation as well
\begin{widetext}
\begin{equation}
  \label{eq:boson_kbe_1}
  [-\partial^{2}_{t_1}-\omega_0^2] D^<(t_1,t_2) = \int_{0}^{t_1}\dd s \Pi^R(t_1,s)D^<(s,t_2) + 
  \int_{0}^{t_2}\dd s \Pi^<(t_1,s)D^A(s,t_2) - \ii \int_{0}^{\beta}\dd \tau \Pi^{\rceil}(t_1,\tau)D^{\lceil}(\tau,t_2)
\end{equation}
\begin{equation}
  \label{eq:boson_kbe_2}
  [-\partial^{2}_{t_2}-\omega_0^2]D^<(t_1,t_2) = \int_{0}^{t_1}\dd s D^R(t_1,s)\Pi^<(s,t_2) + 
  \int_{0}^{t_2}\dd s D^<(t_1,s) \Pi^A(s,t_2) - \ii \int_{0}^{\beta}\dd \tau D^{\rceil}(t_1,\tau)\Pi^{\lceil}(\tau,t_2)
\end{equation}
\end{widetext}
Note those scalar boson KBE have an important difference compared with fermion (or complex boson) KBE where for fermions the time derivatives are $\ii \partial_{t_1} G(t_1,t_2) $ and $-\ii \partial_{t_2} G(t_1,t_2)$. The fermion KBE can be found by replacing the two-point function with those for fermions in the above equations, as well as the free Green's functions. 
 
The second derivative in the boson KBE here seems invalid in the usual subtraction trick~\cite{Karlsson2018, Tuovinen2020}, which needs the first-time derivatives with a relative minus sign as those presented in fermion equations. Here, we may use a workaround that splits the free boson propagator into first derivatives and moves one of them to the right-hand side. We rewrite the boson KBE as
\begin{equation}
  [\ii\partial_{t_1}^{+}+\omega_0][\ii\partial_{t_1}^{+}-\omega_0] D^<(t_1,t_2) = K_1(t_1,t_2)
\end{equation}
\begin{equation}
  [-\ii\partial_{t_2}^{-}+\omega_0][-\ii\partial_{t_2}^{-}-\omega_0] D^<(t_1,t_2) = K_2(t_1,t_2)
\end{equation}
where the $K_1$ and $K_2$ are corresponding terms in \eqref{eq:boson_kbe_2} and \eqref{eq:boson_kbe_1}. Here we also marked $\partial_{t}^{\pm}= \partial_{t} \pm \ii 0^+ $ to remind ourselves they are for retarded (advanced) functions. The above equations can then be rewritten formally as
\begin{equation}
  P_{+}^{-1} \circ P_{-}^{-1} \circ D^< = K_1
\end{equation}
\begin{equation}
       D^< \circ M_{-}^{-1} \circ M_{+}^{-1}= K_2
\end{equation}
where $P_{+}^{-1} \equiv (\ii\partial_{t_1}^{+}+\omega_0)\delta(t_1,t_2)$, $P_{-}^{-1} \equiv (\ii\partial_{t_1}^{+}-\omega_0)\delta(t_1,t_2)$, $M_{+}^{-1} \equiv (\ii\partial_{t_1}^{-}+\omega_0)\delta(t_1,t_2)$, $M_{-}^{-1} \equiv (\ii\partial_{t_1}^{-}-\omega_0)\delta(t_1,t_2)$.
Then we can simply rewrite them by assuming $P_{\pm}^{-1}$ and $M_{\pm}^{-1}$ are invertible in functional sense
\begin{equation}
  \label{eq:boson_kbe_1st_derivative_1}
  (\ii\partial_{t_1}^{+}-\omega_0) \circ D^< = P_{+} \circ K_1
\end{equation}
\begin{equation}
  \label{eq:boson_kbe_1st_derivative_2}
  (-\ii\partial_{t_2}^{-}-\omega_0) \circ D^< =   K_2 \circ M_{+}
\end{equation}
Note, we did the integral by parts to get 
\begin{equation}
  \begin{aligned}
    D(t_1,t_2) \circ M_{-}^{-1} &= D(t_1,t_2) \circ (\ii\partial_{t_1}^{-}-\omega_0)\delta(t_1,t_2) \\
    & =(-\ii\partial_{t_2}^{-}-\omega_0)D(t_1,t_2)
  \end{aligned}
\end{equation}
Now we can subtract \eqref{eq:boson_kbe_1st_derivative_1} from \eqref{eq:boson_kbe_1st_derivative_2} then let $t_1=t_2=t$ resulting a equation looks like master equation~\cite{Karlsson2018,Tuovinen2020}. Notice that $[\omega_0^2,n^{(b)}(t)]=0$ and use $n^{(b)}(t,t)= \ii D^<(t,t)$, and keep in mind the sign difference with fermions, we get
\begin{equation}
  -\partial_{t} n^{(b)}(t) = C_{\mathrm{col}}(t)  
\end{equation}
where we used 
$
  [\partial_{t_1}D(t_1,t_2)+\partial_{t_2}D(t_1,t_2)]_{t_1=t_2=t} = \partial_t D(t,t)
$.
The $C_{\mathrm{col}}(t)$ is 
\begin{equation}
  C_{\mathrm{col}}(t) = [P_{+} \circ K_1](t,t)-[K_2 \circ M_{+}](t,t)
\end{equation}
The $P_{+}$ is retarded function and $M_{+}$ is advanced the solutions are
\begin{equation}
  P_{\pm}(t_1,t_2) = - \ii \theta(t_1-t_2)e^{\pm \ii\omega_0(t_1-t_2)}
\end{equation}
\begin{equation}
  M_{\pm}(t_1,t_2) =  \ii \theta(t_2-t_1)e^{\pm \ii\omega_0(t_1-t_2)}
\end{equation}
One can verify these by $P_{\pm}^{-1} \circ P_{\pm} = \delta(t_1,t_2)$ and the same for $M_{\pm}$.

\subsection{Generalized Kadanoff-Baym ansatz}

A way to approximate the equal time equations is to use the so-called generalized Kadanoff-Baym ansatz~\cite{Lipavsky1986, Karlsson2018, Tuovinen2020}, which says one may approximate the lesser function as
\begin{equation}
  G^<(t_1,t_2) = -G^R(t_1,t_2)\rho(t_2) + \rho(t_1)G^A(t_1,t_2)
\end{equation}
\begin{equation}
  G^>(t_1,t_2) = G^R(t_1,t_2)(1-\rho(t_2))-(1-\rho(t_1))G^A(t_1,t_2)
\end{equation}
where $\rho$ is a single particle density matrix defined for fermions $\rho(t) \equiv -\ii G^{<}(t,t)$. Note the generalized Kadanoff-Bay ansatz usually also approximate the retarded Green's functions to their Hatree-Fock values. However, this approximation is not expected to work well for the strongly correlated phases, both in and out of equilibrium, so we will not apply it to the YSYK here.

\section{Finite-$q$ YSYK Imaginary time}
\label{sec:finite_q_ysyk_imaginary_time}

In equilibrium, we can use the imaginary time formalism, which greatly simplifies many computations. We will focus on half-filling $\mu=0$ here. The lattice contribution is switched off, i.e., $v=0$.

\subsection{Imaginary time effective action}

For the initial condition, we need imaginary time action which should be adapted to our convention~\cite{Aoki2014} $G(\tau_1,\tau_2) = - \braket{\operatorname{T}_{\tau}c(\tau_1)c^{\dagger}(\tau_2)}$
\begin{equation}
  \begin{aligned}
    1 & =\int \dd[G] \prod_{ \tau_1 \tau_2} \delta\left(N G_{ \sigma^{\prime} \sigma}\left(\tau_2, \tau_1 \right) + \sum_i  c_{i \sigma^{\prime} }\left(\tau_2 \right) c_{i \sigma }^{\dagger}(\tau_1)\right) \\
      & =\int \dd[\Sigma,G] \exp \left\{ \sum_{\sigma\sigma'}\int_{\tau_1\tau_2}
    \Sigma_{\sigma\sigma'}(\tau_1,\tau_2) \times \right.\\
    & \qquad \qquad \left.\left(N G_{ \sigma^{\prime} \sigma}\left(\tau_2, \tau_1\right) + \sum_i  c_{i \sigma^{\prime} }\left(\tau_2 \right) c_{i \sigma }^{\dagger}(\tau_1)\right) \right\}
  \end{aligned}
\end{equation}
and the same for $D$ and $\Pi$ but without the spins. Wick rotation $\tau=\ii t$. The imaginary time-effective action can be written as~\cite{Esterlis2019},
\begin{equation}
  \begin{aligned}
    -\frac{1}{N}& S_{\mathrm{eff,im}}= 
    \operatorname{tr} \log \left(G_0^{-1}-\Sigma\right)
    -\frac{\lambda}{2} \operatorname{tr} \log \left(D_0^{-1}-\Pi\right)                                                                        \\
                                      & +  \int_{\tau_1 \tau_2} G\left(\tau_2, \tau_1\right) \Sigma\left(\tau_1, \tau_2\right)                 \\
                                      & -\frac{\lambda}{2} \int_{\tau_1 \tau_2} D\left(\tau_2, \tau_1\right) \Pi\left(\tau_1, \tau_2\right)    \\
                                      & + g^2 \frac{2}{q} \int_{\tau_1 \tau_2} 
                                      % (-1)^{q+1}
                                      \left[G\left(\tau_1, \tau_2\right)\right]^{\frac{q}{2}} \left[- G\left(\tau_2, \tau_1\right)
    \right]^{\frac{q}{2}}
    D\left(\tau_1, \tau_2\right)
  \end{aligned}
\end{equation}
Used $q$ is an even number, and the identities
$
  G_{0}^{-1}
  = (\ii\partial_{-\ii\tau}+\mu)
  = -\partial_{\tau} + \mu
$ and
$
  D_{0}^{-1} = -\partial_{t}^{2} - \omega_0^2
  =-\partial_{-\ii\tau}^{2} - \omega_0^2
  = \partial_{\tau}^{2}- \omega_0^2
$. 
The saddle point equations become
\begin{equation}
  \Sigma = -  g^2 
   R_{q-2}(\tau_1, \tau_2)G\left(\tau_1, \tau_2\right) 
  D(\tau_1,\tau_2)
\end{equation}
\begin{equation}
  \Pi =   2 \lambda^{-1} \frac{2}{q} g^2  
  R_{q-2}(\tau_1, \tau_2)
  G\left(\tau_1, \tau_2\right) G\left(\tau_2, \tau_1 \right)
\end{equation}
where we defined for $q$ an even integer
\begin{equation}
  R_{q-2}(\tau_1, \tau_2) = (-1)^{q-2}\left(-G\left(\tau_1, \tau_2\right) G\left(\tau_2, \tau_1\right)
  \right)^{\frac{q}{2}-1}
\end{equation}
Together with the Dyson equations $
  G^{-1}= G_0^{-1}-\Sigma
$ and
$
  D^{-1}= D_0^{-1}-\Pi
$ we can solve saddle point equations. This is most convenient by using the imaginary time translational invariance $G(\tau_1,\tau_2)=G(\tau_1-\tau_2)$ and $D(\tau_1,\tau_2)=D(\tau_1-\tau_2)$ with imaginary time Fourier transforms $
  G\left(\tau\right)=T \sum_n e^{-i \omega_n \tau} G\left(i \omega_n\right)
$ and
$
  G\left(i \omega_n\right)=\int_0^\beta d \tau e^{i \omega_n \tau} G(\tau)
$, which gives $G_0^{-1}(\ii \omega_n)= \ii \omega_{n} +\mu $, $D_0^{-1}(\ii\nu_n)= (\ii \nu_{n})^2 -\omega_{0}^2 $. 
In numerics, we discretize the imaginary time in $\tau \in [0,\beta)$, and fast Fourier transform the imaginary time to Matsubara frequencies. for the $G(-\tau)$ we use the anti-periodicity of the fermion Green's function $G(-\tau)=-G(\beta-\tau)$. We can verify that $R_{q-2} \geq 0$ by $R_{q-2}(\tau) = [G(\tau)]^{\frac{q}{2}-1}[-G(-\tau)]^{\frac{q}{2}-1} = [G(\tau)G(\beta-\tau)]^{\frac{q}{2}-1} \geq 0$ where $G(\tau)\leq 0 $ if $\tau\in [0,\beta)$. 
% note in \cite{Esterlis2019} (A18) they use D(v)=1/(v^2+w0^2+Pi(v)) which is consistent with us, the only difference is the minus sign. 

\subsection{Compute $q$-dependent exponents for YSYK$_q$}
\label{sec:q_dependent_exponents}

Here we follow~\cite{Esterlis2019} to compute low-frequency exponents of propagators from the self-consistent saddle point equations and introduce the additional parameter $q$ as a variable. Note, we will denote zero temperature Matsubara frequencies as $\omega$ since they are continuous.

\subsubsection{Counting the exponent}
We first omit the prefactor and focus only on exponents for a schematic analysis. We assume a $q$ dependent ansatz for the imaginary time fermion propagator
$
G(\tau) \sim a |\tau|^{\frac{-4\Delta}{q}}
$
which is similar to the usual SYK~\cite{Maldacena2016}. The boson self-energy is then
$
  \Pi(\tau) \sim \frac{ g^2}{\lambda} G(\tau)^{\frac{q}{2}}G(-\tau)^{\frac{q}{2}} \simeq \frac{a^{q} g^2}{\lambda}|\tau|^{-4\Delta}
$
we keep the prefactor for couplings $g$ and $\lambda$ and omit others. Fourier it to frequency space and keep only the finite contribution 
$\Pi(\omega) \sim \frac{a^{q} g^2}{\lambda}|\omega|^{4\Delta-1}$. For low frequencies 
$D(\omega) \sim \frac{1}{\Pi(\omega)}$, thus 
$D(\tau) \sim \frac{\lambda}{a^{q} g^2}|\tau|^{4\Delta-2}$. The fermion self-energy then follows 
$\Sigma(\tau)\sim g^2 G(\tau)^{\frac{q}{2}}G(-\tau)^{\frac{q}{2}-1}D(\tau)\sim \frac{\lambda}{a^q} |\tau|^{\frac{4\Delta}{q}-2}$. Fourier again 
$\Sigma(\omega) \sim \frac{\lambda}{a}|\omega|^{1-\frac{4\Delta}{q}}$. $G(\omega)\sim \frac{1}{\Sigma(\omega)}\sim \frac{a}{\lambda}|\omega|^{\frac{4\Delta}{q}-1}$ which means 
$G(\tau) \sim \frac{a}{\lambda}|\tau|^{\frac{-4\Delta}{q}}$ and we complete one consistent loop.

\subsubsection{Compute the exponent}
The numerical value of the exponent $\Delta$ can be fixed by taking into account the exact prefactor~\cite{Maldacena2016, Esterlis2019, Wang2020}. Note we use Fourier convention $1 \overset{\mathrm{F.T.}}{\longrightarrow} 2\pi \delta(\omega)$. The useful Fourier transform identities are
\begin{equation}
  \int \dd t e^{-\ii \omega t}\operatorname{sgn}(t)|t|^{x-1} =-2 \ii  \sin \left(\frac{\pi  x}{2}\right) \Gamma (x) \operatorname{sgn}(\omega ) | \omega | ^{-x}
\end{equation}
\begin{equation}
  \int \frac{\dd \omega}{2\pi}  e^{\ii \omega t}\operatorname{sgn}(\omega)|\omega|^{x-1} =\frac{i}{\pi}  \sin \left(\frac{\pi  x}{2}\right) \Gamma (x) \operatorname{sgn}(t)| t| ^{-x}
\end{equation}
Here $x$ is a real number, and the convergence of Fourier transforms requires $0<x<2$.

With the help of the above identities, we can compute the $\Delta$ by going through one self-consistent loop. Starting from the ansatz
\begin{equation}
  \label{eq:G_conformal}
  G(\tau) = a \operatorname{sgn}(\tau)|\tau|^{\frac{-4\Delta}{q}}
\end{equation}
where $q$ is an even integer and $\Delta$ is a real number. Note the prefactor $a$ is a non-universal factor and its exact value depends on UV details~\cite{Wang2020}. Approximations can be made for the $a$ from scaling solutions, see~\cite{Esterlis2019, Wang2020a}. The finite part of the boson self-energy can be written down by plugging in~\eqref{eq:G_conformal}
\begin{equation}
  \begin{aligned}
    \Pi(\tau) 
    &=  \frac{2 g^2}{\lambda} \frac{2}{q} (-1)^{q-2}G(\tau)^{\frac{q}{2}}G(-\tau)^{\frac{q}{2}} \\
    &= \frac{2 g^2}{\lambda } \left(a \operatorname{sgn}(\tau ) | \tau | ^{-\frac{4 \Delta }{q}}\right)^q
  \end{aligned}
\end{equation}
use $q$ an even number, we have $(-1)^{q-2}=1$ and $\operatorname{sgn}(\tau)^q=1$. Then the Fourier of $\Pi(\tau)$ is
\begin{equation}
  \Pi(\ii \omega) = \frac{4 g^2}{\lambda } \frac{2}{q} a^q \sin (2 \pi  \Delta ) \Gamma (1-4 \Delta ) | \omega | ^{4 \Delta -1}
\end{equation}
with $4\Delta-1>0$ for a well-defined Fourier transform. Use $D(\ii\omega)=\frac{1}{-\Pi(\ii\omega)}$ and Fourier the time-space boson propagator reads
\begin{equation}
  \label{eq:conformal_D_tau}
  D(\tau) = \frac{\lambda}{4 \pi  g^2}\frac{q}{2} (1-4 \Delta )    a^{-q} \cot (2 \pi  \Delta ) | \tau | ^{-(2-4 \Delta)}
\end{equation}
Here we require $2-4 \Delta>0$ for a convergent Fourier transform. The fermionic self-energy is then
\begin{equation}
  \begin{aligned}
    \Sigma(\tau) &= -g^2  (-1)^{q-2} G(\tau)^{\frac{q}{2}}G(-\tau)^{\frac{q}{2}-1}D(\tau) \\
    % &= \frac{\lambda}{4 \pi }(1-4 \Delta )   a^{-q} \cot (2 \pi  \Delta ) | \tau | ^{4 \Delta -2} \left(a \operatorname{sgn}(\tau ) | \tau | ^{-\frac{4 \Delta }{q}}\right)^{q-1}\\
    &=\frac{\lambda}{4 \pi } \frac{q}{2} (1-4 \Delta )   a^{-1} \cot (2 \pi  \Delta ) \operatorname{sgn}(\tau ) | \tau | ^{\frac{4 \Delta}{q} -2} 
  \end{aligned}
\end{equation}
Fourier transform $\Sigma(\tau)$ back to frequency space
\begin{equation}
  \begin{aligned}
    \Sigma(\ii \omega) =& \frac{\ii \lambda}{2 \pi  a} \frac{q}{2} (1-4 \Delta ) \cot (2 \pi  \Delta ) \cos \left(\frac{2 \pi  \Delta }{q}\right) \times \\
    &\Gamma \left(\frac{4 \Delta }{q}-1\right) \operatorname{sgn}(\omega)| \omega | ^{1-\frac{4 \Delta }{q}}
  \end{aligned}
\end{equation}
Use $G_{1}(\ii\omega)=\frac{1}{-\Sigma(\ii\omega)}$, we get
\begin{equation}
  \begin{aligned}
    G_{1}(\tau) =   &\frac{-4 a \tan (2 \pi  \Delta ) (q-4 \Delta )  \tan \left(\frac{2 \pi  \Delta }{q}\right) }{(4 \Delta -1) \lambda  q^2} \times\\
    &\operatorname{sgn}(\tau )| \tau | ^{-\frac{4 \Delta }{q}}
  \end{aligned}
\end{equation}
Apply the consistency condition
$
  G(\tau) \overset{!}{=} G_1(\tau)
$
which implies
\begin{equation}
  \label{eq:finite_q_exponent}
  1 \overset{!}{=} - \frac{4 \tan (2 \pi  \Delta ) (q-4 \Delta ) \tan \left(\frac{2 \pi  \Delta }{q}\right)}{(4 \Delta -1) \lambda  q^2}
\end{equation}
Here the $\Delta$ should satisfy $\frac{1}{4}<\Delta<\frac{1}{2}$ for a well-defined Fourier transform. We can compare this result with the known exponent~\cite{Esterlis2019} for $q=2$ and $\lambda=1$, and they agree.
% the exponent is $\Delta \simeq 0.42037$. 
% The above equation gives
% $
%   \frac{2 (2 \Delta -1) \tan (\pi  \Delta ) \tan (2 \pi  \Delta )}{4 \Delta -1}=1
% $. Note that $\tan (\pi  \Delta ) \tan (2 \pi  \Delta )=(\sec (2 \pi  \Delta )-1)$.

Several limits can be taken to simplify the \ref{eq:finite_q_exponent}. If we assume $\Delta$ is not $q$ dependent, then for $q\gg 1$ we can expand around $\frac{1}{q} \to 0$ where we have  up to first order
\begin{equation}
  1 \overset{!}{=}-\frac{2  (\pi  \Delta  \tan (2 \pi  \Delta ))}{(4 \Delta -1)  } \frac{1}{ \lambda_1}+O\left(\frac{1}{q^3}\right)
\end{equation}
where we defined
\begin{equation}
  \lambda = \frac{4 \lambda_1}{q^2}
\end{equation}
The factor $4$ is to cancel $q^2$ when $q=2$. We can solve $\Delta$ for $q=\infty$ and $\lambda_1=1$, the solution is $\Delta \simeq 0.455488$, which is not saturating to $\frac{1}{2}$ or $\frac{1}{4}$. For large (small) $\lambda_1$ the $\Delta$ goes to its lower (upper) bounds $\lambda_1\to 0$, $\Delta=\frac{1}{2}$ and $\lambda_1\to \infty$, $\Delta=\frac{1}{4}$, which we also checked numerically.

\begin{figure}
  \centering
  \includegraphics[width=0.8\linewidth]{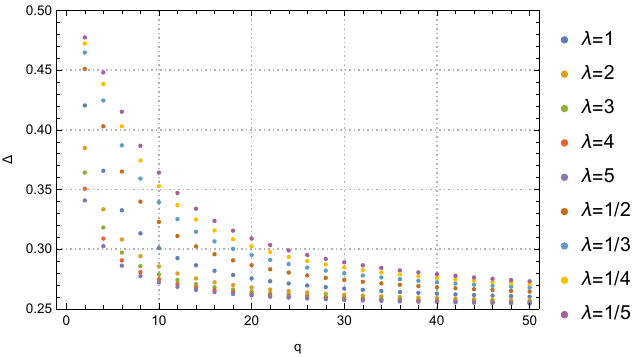}
  \caption{The exponent $\Delta$ by numerically solve the Eq.\eqref{eq:finite_q_exponent}.
  }
\label{fig:finite_q_exponent_1}
\end{figure}

\begin{figure}
  \centering
  \includegraphics[width=0.45\linewidth]{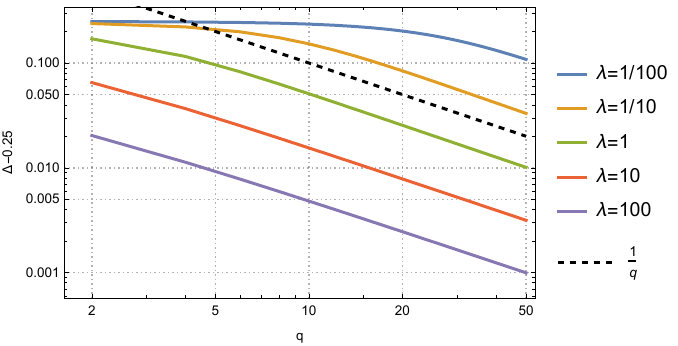}
  \includegraphics[width=0.4\linewidth]{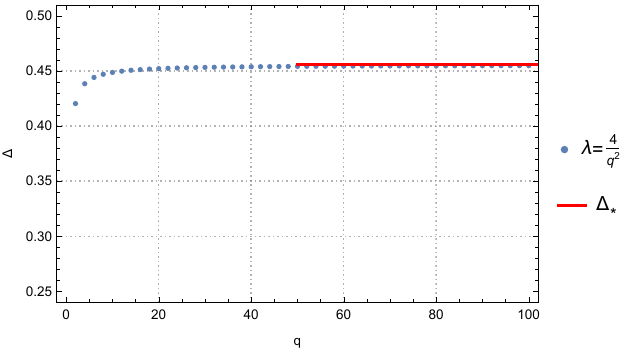}
  \includegraphics[width=0.45\linewidth]{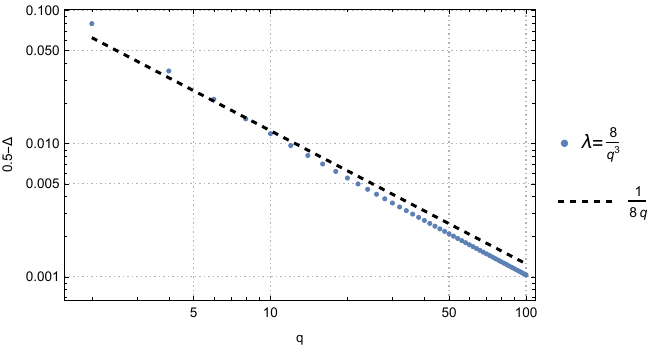}
  \includegraphics[width=0.45\linewidth]{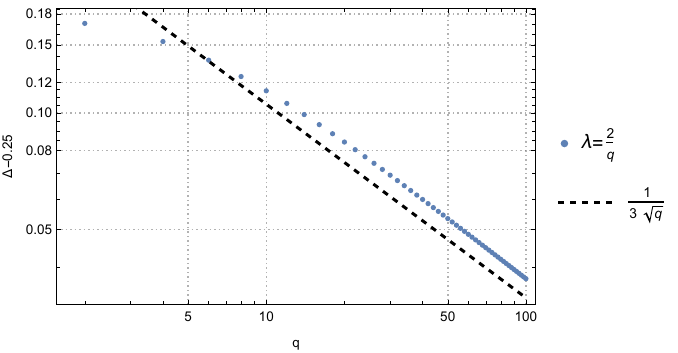}
  \caption{Numerical solutions of the exponent $\Delta$. 
  (Upper left) The log-log plot $(\Delta-\frac{1}{4})$ vs. $q$ with different $\lambda$. The points are joined for better visibility.
  (Upper right) Choosing a $q$ dependent $\lambda=\frac{4}{q^2}$. This gives an exponent other than $\frac{1}{2}$ or $\frac{1}{4}$ for large $q$. The red line marks the $\lambda$-$q$ double scaling limit $\frac{1}{q}\to 0$ with $\lambda=\frac{4}{q^2}$, the value is $\Delta_{*}\simeq 0.45548817958394894$.
  (Lower left) The log-log plot $(\frac{1}{2}-\Delta)$ vs. $q$ with $\lambda=\frac{8}{q^3}$. One can see that $\Delta$ approaching $1/2$ as a linear function of $\frac{1}{q}$.
  (Lower right) The log-log plot $(\Delta-\frac{1}{4})$ vs. $q$ with $\lambda=\frac{2}{q}$, which can cancel the explicit prefactor in front of boson self-energy. In this case $\Delta$ is approaching $\frac{1}{4}$ algebraically as a linear function of $\frac{1}{\sqrt{q}}$.
  }
\label{fig:finite_q_exponent_2}
\end{figure}

We show numerical solutions of the \eqref{eq:finite_q_exponent} in Fig.~\ref{fig:finite_q_exponent_1} and Fig.~\ref{fig:finite_q_exponent_2}. Several values of $\Delta$ are summarized in Table \ref{tab:exponent_table1} for fixed $\lambda$ and can be compared to numerical solutions of the saddle point equations.

\subsection{The $q$ dependent propagators at finite temperature}
\label{sec:q_dependent_propagators_eq}

The finite temperature fermion and boson propagators in the scaling limit can be written down now from previous computations. The time reparameterization symmetry allows us to map a zero temperature solution to an imaginary time circle~\cite{Sachdev2015, Maldacena2016, Davis2023}. Here we extend those results for general $q$ but restrict ourselves to spectral symmetry, i.e., half-filling. The results can be obtained straightforwardly by noticing that $G(\tau) \sim |\tau|^{\frac{-4\Delta}{q}}$, and the formal boson scaling dimension is not changed $D(\tau)\sim |\tau|^{2-4\Delta}$. The two-point function has the symmetry
\begin{equation}
  \begin{aligned}
    G\left(\tau, \tau^{\prime}\right) & \rightarrow\left[f^{\prime}(\tau) f^{\prime}\left(\tau^{\prime}\right)\right]^{\frac{4\Delta}{q}} \frac{l(\tau)}{l\left(\tau^{\prime}\right)} G\left(f(\tau), f\left(\tau^{\prime}\right)\right), \\
    \Sigma\left(\tau, \tau^{\prime}\right) & \rightarrow\left[f^{\prime}(\tau) f^{\prime}\left(\tau^{\prime}\right)\right]^{1-\frac{4\Delta}{q}} \frac{l(\tau)}{l\left(\tau^{\prime}\right)} \Sigma\left(f(\tau), f\left(\tau^{\prime}\right)\right) \\
    D\left(\tau, \tau^{\prime}\right) & \rightarrow\left[f^{\prime}(\tau) f^{\prime}\left(\tau^{\prime}\right)\right]^{1-2 \Delta} D\left(f(\tau), f\left(\tau^{\prime}\right)\right), \\
    \Pi\left(\tau, \tau^{\prime}\right) & \rightarrow\left[f^{\prime}(\tau) f^{\prime}\left(\tau^{\prime}\right)\right]^{2 \Delta} \Pi\left(f(\tau), f\left(\tau^{\prime}\right)\right)
  \end{aligned}
\end{equation}
where $l(\tau)$ is a arbitrary function from the emergent $U(1)$ symmetry of complex fermions~\cite{Sachdev2015} and at half-filling it can be chosen trivially $l(\tau)=1$. Likewise $f$ is arbitrary and $f'(x)\equiv \dd f(x)/\dd x$. To map the time to the finite temperature circle we can choose $f(\tau)=\tan \left(\frac{\pi \tau}{\beta}\right)$ which maps $\frac{1}{|\tau|} \to \frac{\pi}{\beta |\sin \frac{\pi \tau}{\beta}|}$. Use $D(\tau_1,\tau_2)=D(\tau_1-\tau_2)$ and same for $G$ we get
\begin{equation}
G(\tau)=  A_G  
  \operatorname{sgn}(\tau)
\left(\frac{\pi}{\beta\left|\sin \frac{\pi \tau}{\beta}\right|}\right)^{\frac{4\Delta}{q}}\\
\end{equation}
and 
\begin{equation}
  D\left(\tau\right)=A_D\left(\frac{\pi}{\beta\left|\sin \frac{\pi \tau}{\beta}\right|}\right)^{2-4\Delta}
\end{equation}
$A_G$ and $A_D$ are overall constants.

%%%%%%%%%%%%%%%%%%%%%%%%%%%%%%%%%%%%%%%
%% bibliography
%%%%%%%%%%%%%%%%%%%%%%%%%%%%%%%%%%%%%%%
% \clearpage
% \twocolumngrid
% \nocite{*}
% \bibliography{apssamp}% Produces the bibliography via BibTeX.
\bibliography{YSYK_latt_q_np}
\end{document}